\documentclass[a4paper,fleqn,usenatbib]{mnras}

\usepackage{newtxtext,newtxmath}

\usepackage[T1]{fontenc}
\usepackage{ae,aecompl}

\usepackage{graphicx}	
\usepackage{amsmath}	
\usepackage{amssymb}	

\title[Testing the two-corona model on NGC 4593]{High-energy monitoring of NGC 4593 II.
	\\ Broadband spectral analysis: testing the two-corona model}

\author[R. Middei et al.]{
R. Middei,$^{1}$\thanks{E-mail: riccardo.middei@uniroma3.it}
S. Bianchi,$^{1}$ P.-O. Petrucci,$^{2}$
F. Ursini,$^{3}$ 
M. Cappi,$^{3}$ B. De Marco,$^{4}$\newauthor A. De Rosa,$^{5}$ J. Malzac,$^{6}$  A. Marinucci, $^{1}$ G. Matt, $^{1}$ G. Ponti,$^{7}$ and A. Tortosa $^{5}$
\\
$^{1}$Dipartimento di Matematica e Fisica, Universit\`a degli Studi Roma Tre, via della Vasca Navale 84, I-00146 Roma, Italy\\
$^{2}$Univ. Grenoble Alpes, CNRS, IPAG, F-38000 Grenoble, France\\
$^{3}$INAF-Osservatorio di astrofisica e scienza dello spazio di Bologna, Via Piero Gobetti 93/3, 40129 Bologna, Italy.\\
$^{4}$Nicolaus Copernicus Astronomical Center, PL-00-716 Warsaw, Poland.\\
$^{5}$INAF/Istituto di Astrofisica e Planetologia Spaziali, via Fosso del Cavaliere, 00133 Roma, Italy.\\
$^{6}$IRAP, Universit\'e de Toulouse, CNRS, UPS, CNES, Toulouse, France.\\
$^{7}$Max-Planck-Institut fur extraterrestrische Physik, Giessenbachstrasse, D-85748 Garching, Germany.\\
}

\date{Accepted XXX. Received YYY; in original form ZZZ}

\pubyear{2015}

\begin{document}
\label{firstpage}
\pagerange{\pageref{firstpage}--\pageref{lastpage}}
\maketitle

\begin{abstract}
It is widely believed that the primary X-ray emission of AGN is due to the Comptonisation of optical-UV photons from a hot electron corona, while the origin of the 'soft-excess' is still uncertain and matter of debate. A second Comptonisation component, called warm corona, was therefore proposed to account for the soft-excess, and found in agreement with the optical-UV to X-ray emission of a sample of Seyfert galaxies. In this context, we exploit the broadband \textit{XMM-Newton} and \textit{NuSTAR} simultaneous observations of the Seyfert galaxy NGC 4593 to further test the so called "two corona model". The NGC 4593 spectra are well reproduced by the model, from the optical/UV to the hard X-rays. Moreover, the data reveal a significant correlation between the hot and the warm corona parameters during our monitoring campaign.
\end{abstract}




\section{Introduction}

\begin{figure}
	\includegraphics[width=0.5\textwidth]{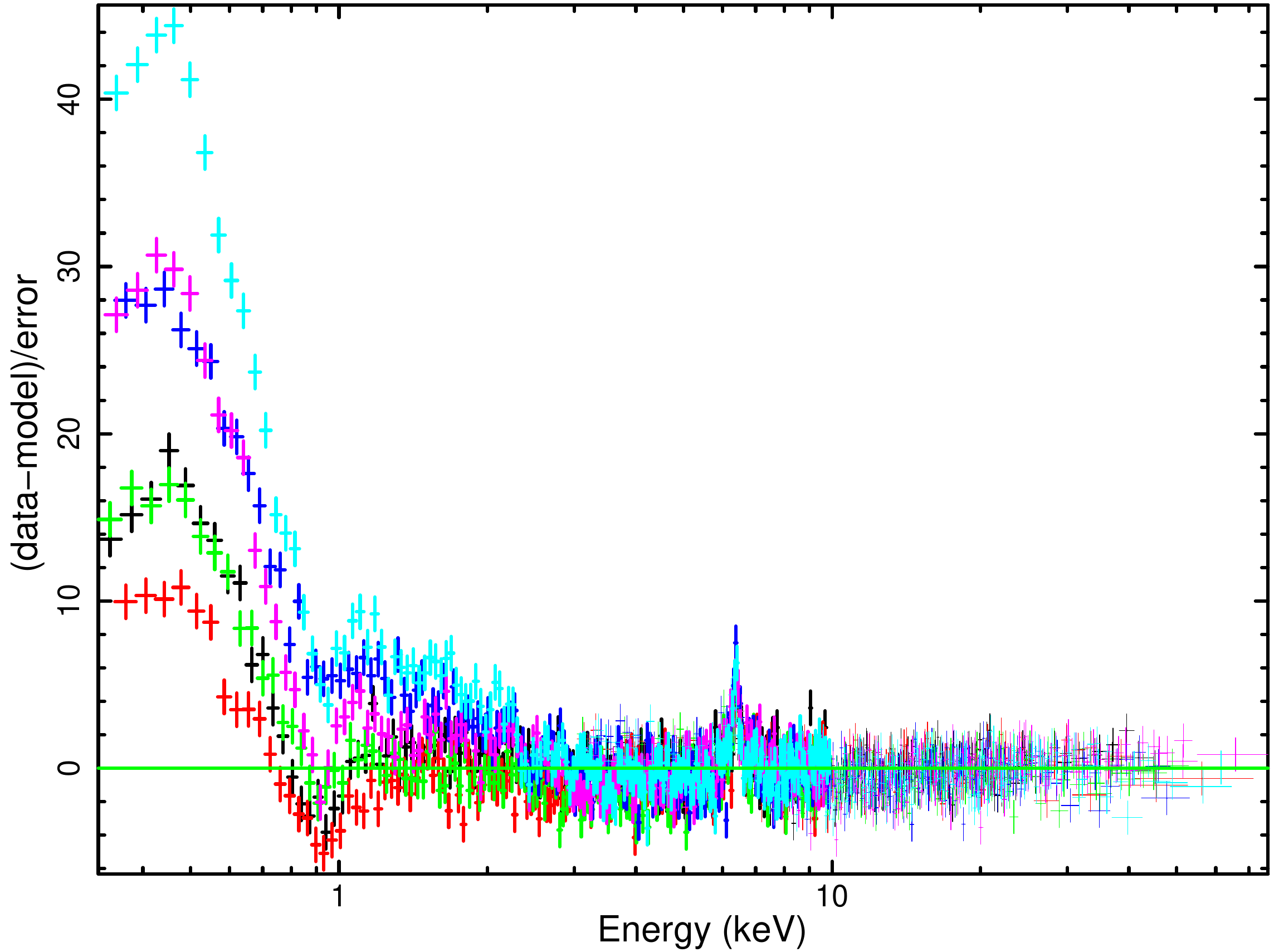}
	\caption{\small{Residuals of the \textit{XMM-Newton} and \textit{NuSTAR} data to a power model fitted in the 4-10 keV energy range and extrapolated down to 0.3 keV. A prominent and variable soft-excess is clearly present below 2-3 keV. Black, red, green, blue, magenta and cyan colours refer to obs. 1a, obs. 1b, obs. 2, obs. 3, obs. 4 and obs. 5, respectively. This colour code is adopted in the whole paper}}\label{soft_excess}
\end{figure}

According to the standard paradigm, the intrinsic X-ray emission of active galactic nuclei (AGN) is the result of an inverse-Compton process in which seed optical-UV photons arising from the accretion disc interact with a compact distribution of thermal electrons, the hot corona \cite[e.g.][]{Haar91,haar93,haar94}. 
This mechanism accounts for the observed power law like X-ray continuum. The optical depth ($\tau_{\rm{hc}}$) and the temperature of the hot coronal electrons (kT$_{\rm{hc}}$) affect the spectral shape of the primary continuum. On the other hand, since the energy gain of the up-scattering disc photons is limited, a high energy cut-off is expected \cite{Rybi79}. This high energy cut-off can be therefore interpreted as the signature of thermal Comptonisation itself, thus many efforts were spent to measure this spectral feature \cite[e.g.][]{Pero00,Nica00,DeRo02,dadina07,Moli09,Molina13,Malizia14}.
More recently, thanks to its unprecedented sensitivity above 10 keV, \textit{NuSTAR} allowed for a large number of cut-off measurements \cite[e.g.][]{Fabi15,Fabi17,Tort18a}.\\
\indent On the other hand, in the soft X-rays, an excess of photons rising below 1 keV above the extrapolated high energy emission is commonly observed in a large percentage of AGN \cite[e.g.][]{Walt93,Page04,Gier04,Ponti06,Crum06,Bian09}. This spectral feature, also called soft-excess, has been the object of many speculations \cite[e.g.][]{Done12} and, at present, its origin is still a matter of debate. Different models accounting for blurred  ionised  reflection, ionised absorption or partial covering, and thermal Comptonisation have been proposed \cite[e.g.][]{Magd98,Crum06,Done07,Jin12,Done12}.\\
\indent In this context, \cite{Petr13} studying the rich data set of the \textit{XMM-Newton}-\textit{INTEGRAL} multiwavelength campaign \cite[][]{Kaas11} found that the spectrum of Mrk 509 can be described by a two-corona model: beside a hot and optically thin corona accounting for the hard X-rays, a warm ($kT$$\sim$1 keV) and optically thick ($\tau\sim15$) corona was in fact able to reproduce the Mrk 509 optical-UV emission and its soft-excess. The two-corona model was then adopted in analysing the high signal-to-noise (S/N) data belonging to the multiwavelength campaign on NGC 7469 \cite[][]{Middei18}. \cite{Porq18}, working on high quality data of Ark 120, also found that the source emission is dominated by two temperature Comptonisation processes. Finally, the two-corona model was successfully tested on a larger sample of Seyfert galaxies \cite[][]{pop18}, though the energy coupling between the two coronae and the outer disc also suggests that part of the optical-UV flux could be produced by the outer standard disc \cite[][]{Kubota18}.\\
\indent The multiwavelength campaign on NGC 4593 provides a further excellent data-set for testing the two-corona model. The source was observed simultaneously by \textit{XMM-Newton} and \textit{NuSTAR}, thus data extend from the optical-UV band up to the hard X-rays.
A phenomenological spectral analysis of the campaign is reported in \cite{Ursi16}, hereafter P1. In P1 the source was found to be variable, both in flux and spectral shape, and the characteristic softer-when-brighter behaviour was observed.
During the observational campaign, a strongly variable high energy cut-off was measured (E$_{\rm{cut-off}}$ from 90$^{+40}_{-20} $ keV to $>$700 keV), and the spectral index varied on timescales down to two days between 1.6 and 1.8.
A prominent \ion{Fe} K$\alpha$ line was measured, best explained as the superposition of a narrow constant component and a broader component likely arising, respectively, from distant cold material and from cirmunuclear matter at about $\simeq$40 R$_{\rm{grav}}$ (R$_{\rm{grav}}$=2GM/c$^2$). In agreement with past studies on this source, a warm absorber consistent with a two-phase ionized outflow was needed to reproduce the data set.
From the analysis of \textit{RGS} data, an additional photoionised component in emission was required to fit the data. Furthermore, NGC 4593 showed a remarkable and variable soft-excess during the monitoring, see Fig.~\ref{soft_excess}.
In this work, exploiting the high S/N data of the NGC 4593 observational campaign, we test Comptonisation, and, in particular, the two-corona model on this source.
The standard cosmology \textit{$\Lambda$CDM} with H$_0$=70 km/s/Mpc, $\Omega_m$=0.27, $\Omega_\lambda$=0.73, is adopted.

\section{Data}	\begin{table}
	\centering
	\caption{\small{For each satellite the observation ID, the start date and the net exposure time are reported.}}
	\begin{tabular}{c c c c}
		\hline
		Obs. satellite & Obs. ID & Start date & Net exp. \\
		& & yyyy-mm-dd & ks \\
		\hline
		\hline
		\textit{XMM-Newton} & 0740920201 & 2014-12-29 & 16\\
		\textit{NuSTAR} & 60001149002 &  & 22 \\
		\hline
		\textit{XMM-Newton} & 0740920301 & 2014-12-31 & 17\\
		\textit{NuSTAR} & 60001149004 &  & 22 \\
		\hline
		\textit{XMM-Newton} & 0740920401 &2015-01-02& 17 \\
		\textit{NuSTAR} & 60001149006 && 21 \\
		\hline
		\textit{XMM-Newton} &0740920501  &2015-01-04 & 15 \\
		\textit{NuSTAR} & 60001149008 && 23 \\
		\hline
		\textit{XMM-Newton} & 0740920601 & 2015-01-06 & 21 \\
		\textit{NuSTAR} & 60001149010&  & 21 \\
		\hline
	\end{tabular}
\end{table}

\begin{figure}
	\includegraphics[width=0.45\textwidth]{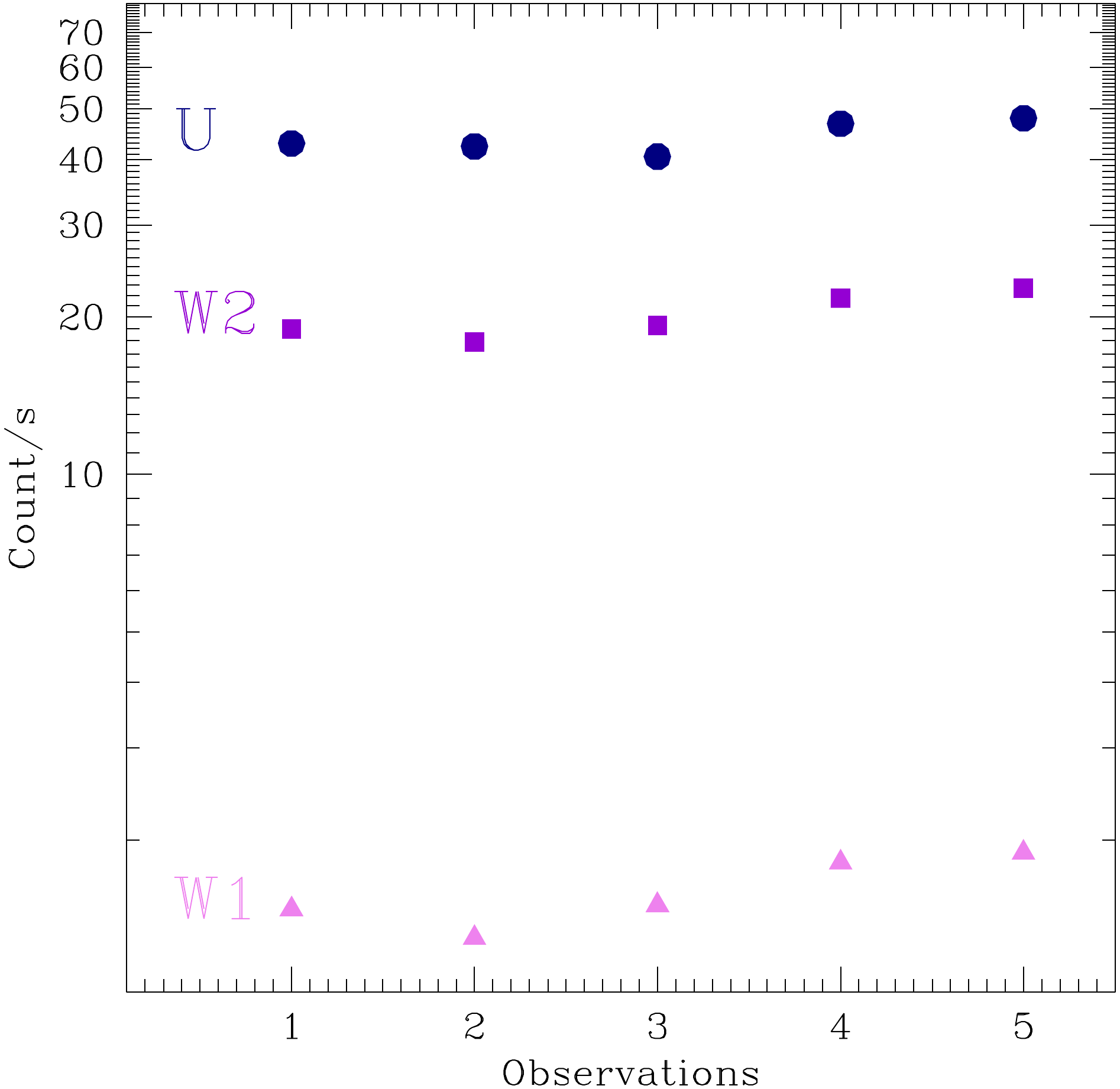}
	\caption{\small{The \textit{OM} rates for the available filters are shown.}\label{om}}
\end{figure}

\begin{figure*}
	\includegraphics[]{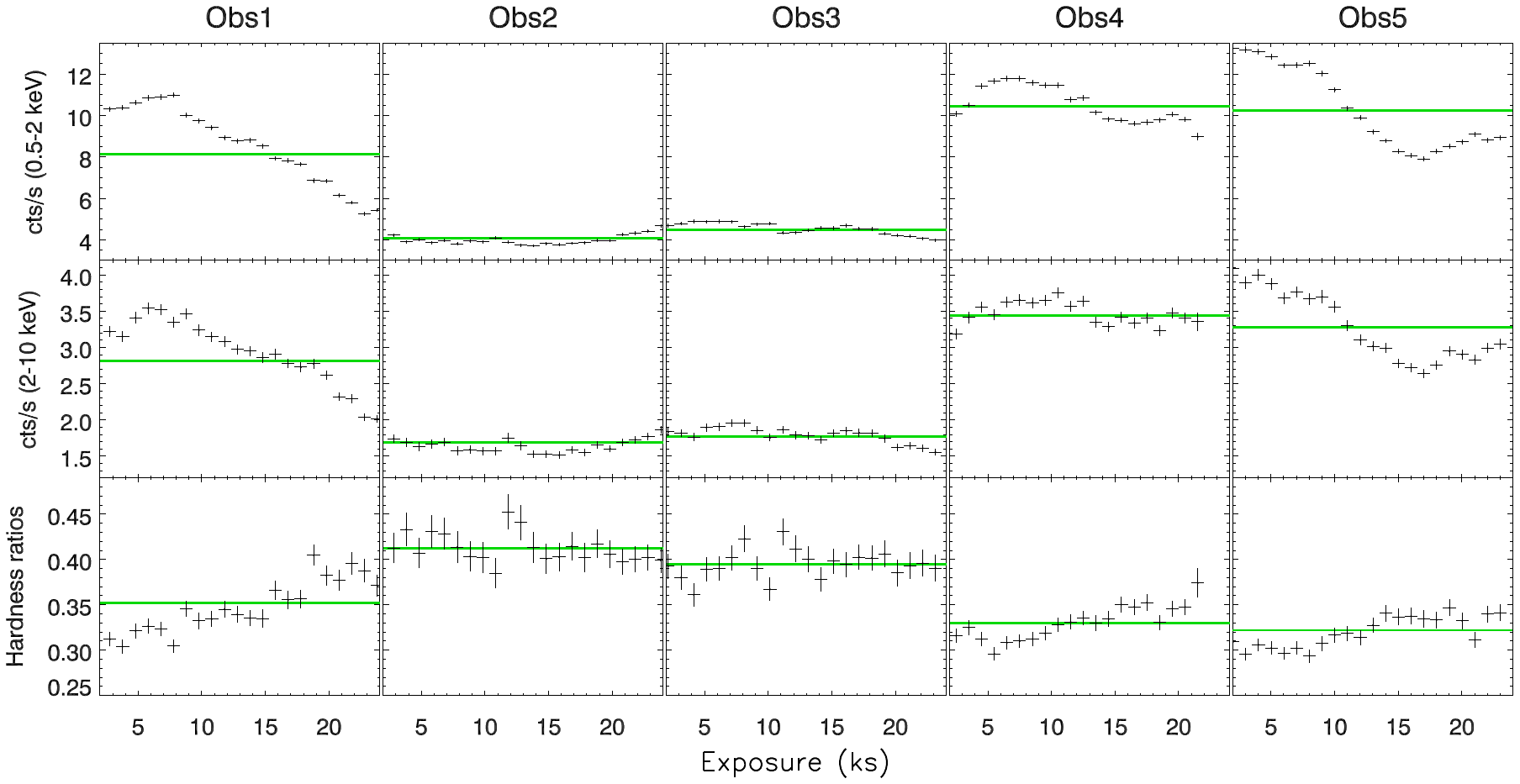}
	\caption{\small{The \textit{XMM-Newton} light curves in the 0.5-2 keV and 2-10 keV energy bands are shown in the top and middle panels respectively. Bottom panels display the ratios between the hard light curves and those computed in the soft band. The adopted time binning is 1 ks for all the observations, and the solid green lines account for the average count rates within each pointing.}\label{lc}}
\end{figure*}

The data set analysed here belongs to the joint \textit{XMM-Newton} and \textit{NuSTAR} monitoring program on NGC 4593, and consists of 5$\times\sim$20 ks simultaneous observations. The monitoring covers the time period between December 29 2014 and January 06 2015, with consecutive pointings being about two days apart, (see Tab. 1).\\
\indent\textit{XMM-Newton} data of NGC 4593 were obtained with the EPIC cameras \cite[][]{stru01,Turn01} in \textit{Small Window} mode with the medium filter applied. Because of its larger effective area with respect to the two MOS cameras, we only used the results for the PN instrument.
Data are processed using the \textit{XMM-Newton} Science Analysis System ($SAS$, Version 16.1.0). The choice of source extraction radius and the screening for high background time intervals are performed by an iterative process that maximizes the S/N, as in \cite{Pico04}. The source radii span between 20 and 40 arcsec. The background is then extracted in a blank region close to the source using a region with a radius of 50 arcsec. We then rebbined all the spectra to have at least 30 counts for each bin, and without oversampling the instrumental energy resolution by a factor larger than 3. Moreover, for the present analysis we take advantage of data provided by the Optical Monitor, \cite[\textit{OM},][]{Maso01}, on-board \textit{XMM-Newton}. This instrument observed NGC 4593 with the filters U (3440 \AA), UVW1 (2910 \AA) , UVW2 (2120 \AA) for all the pointings of the campaign. Data provided by the \textit{OM} are extracted using the on-the-fly data processing and analysis tool \textit{RISA}, the Remote Interface SAS Analysis. To convert the spectral points into convenient format to be analysed with \textit{Xspec} \cite[][]{Arna96}, we used the standard task \textit{om2pha}.
The light curves for the \textit{OM} various filters are reported in Fig~\ref{om}.\\
\indent \textit{NuSTAR} data were reduced taking advantage of the standard pipeline ($nupipeline$) in the \textit{NuSTAR} Data Analysis Software (nustardas release: nustardas\_14Apr16\_v1.6.0, part of the \textit{heasoft} distribution). The adopted calibration database is 20171204. High scientific products were then obtained using the standard \textit{nuproducts} routine for both the hard X-ray detectors \textit{FPMA/B} carried on the \textit{NuSTAR} focal plane. A circular region with a  radius  of  75  arcsec is used to extract the source spectra, while the background is extracted from a blank area with the same radius, close to the source. We have then binned the \textit{NuSTAR} spectra in order to have a S/N greater than 5 in each spectral channel, and to not oversample the resolution by a factor greater than 2.5.
\\
\indent Light curves and hardness ratios are discussed and shown in P1 to which we refer the reader. 
In all the observations of the campaign both spectral and flux variability can be observed. In particular, in the first pointing variability is very significant (see Fig.~\ref{lc}), thus we split observation one in two segments of 10 ks each. Then similarly to P1, we are left with six spectra for testing the two-corona model.\\
\indent Finally, we notice that in other multiwavelength campaigns different photon index estimates were obtained from \textit{XMM-Newton} and \textit{NuSTAR} data in AGN and X-ray binaries \cite[e.g.][]{Capp16,Middei18,Ponti18}. This issue, likely due to residual-intercalibration, leads to a variable $\Delta\Gamma_{\rm{\textit{XMM-NuSTAR}}}$, and, usually, $\Gamma_{\rm{\textit{NuSTAR}}}$ is steeper with respect to $\Gamma_{\rm{\textit{XMM-Newton}}}$. The present data set is marginally affected by this intercalibration problem since only in observation two we find a discrepancy among the photon index estimates: $\Delta\Gamma_{\rm{\textit{XMM-NuSTAR}}}$=0.07. For the sake of simplicity, in Tab. 2  we quoted the photon index derived by \textit{NuSTAR}. Furthermore, we notice that allowing for different values of photon index does not modify the quality of the fit for the present data set.

\section{Spectral Analysis}

We performed the spectral analysis using \textit{Xspec}. 
In the forthcoming fits a free cross-calibration constant is used to take into account residual calibration problems between \textit{XMM-Newon} and \textit{NuSTAR} data. The \textit{NuSTAR} modules are in good agreement with each other ($\sim$2\%), and the detectors of both satellites agree within $\lesssim$10$\%$. 
In this analysis we take into account the absorption due to the Galactic hydrogen column through the \textit{Xspec} model \textit{phabs} for which the Galactic column density is set to N$_{\rm{H}}$=1.89$\times$10$^{20}$ cm$^{-2}$ \cite[][]{Kalb05}. During the fitting procedure, the Galactic N$_{\rm{H}}$ is kept fixed. 

\begin{figure}
	\includegraphics[width=0.5\textwidth]{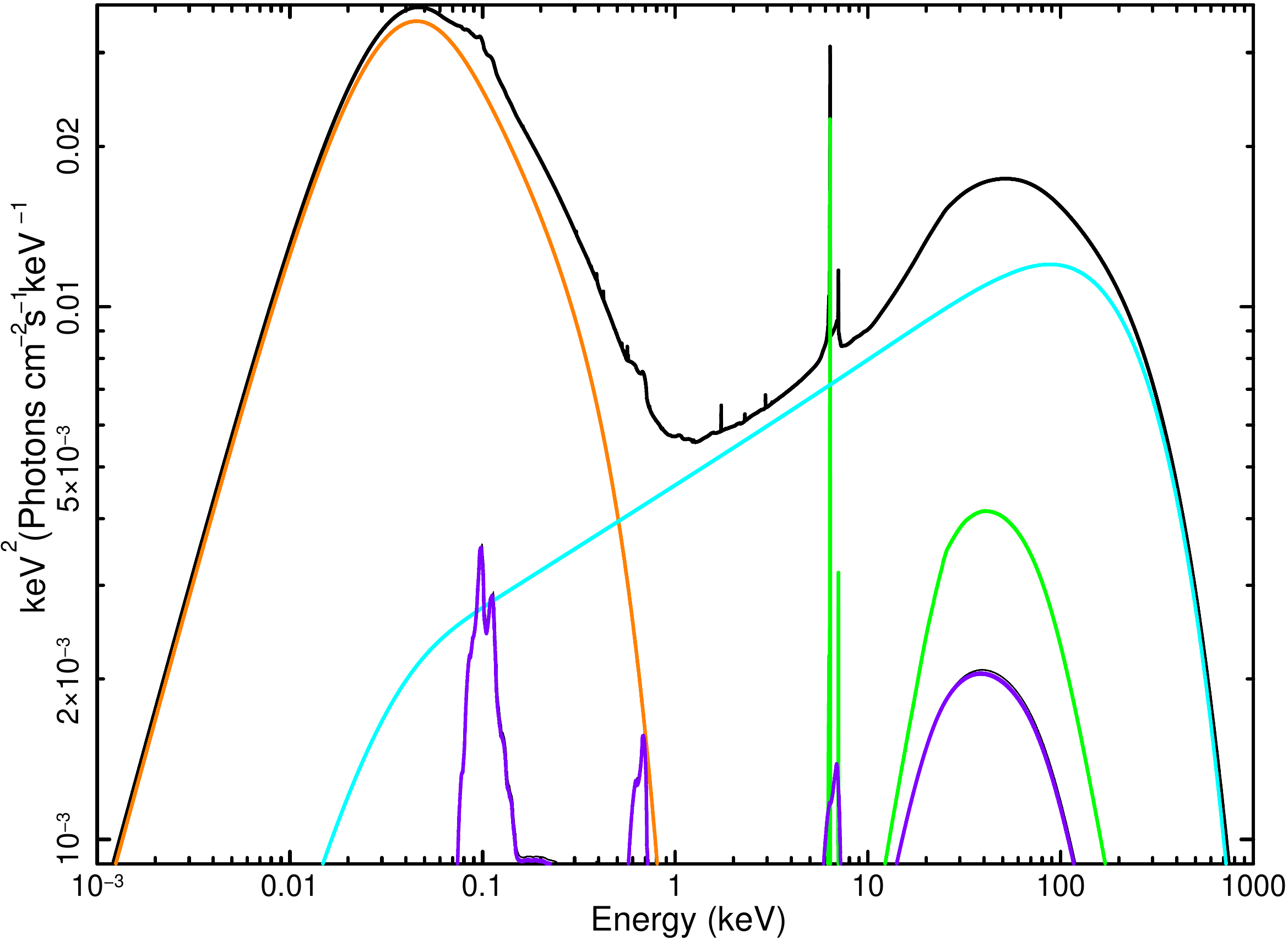}
	\caption{\small{The Comptonising and reflection components are displayed. In orange and cyan the warm and hot corona contribution while components in green and purple account for distant and relativistic reflection, respectively. For the sake of simplicity we do not report the different absorbing components. Example of the Comptonising and reflection components from the best fits of Obs 1b.}}\label{best_mo}
\end{figure}

\indent\textbf{Optical-UV}: This spectral investigation extends from the optical-UV domain up to hard X-rays, thus we have to consider the Broad Line Region (BLR) contribution. In fact, the BLR is responsible for the so-called small blue bump (SBB) at about 3000 \AA. We account for this component using an additive table in \textit{Xspec}. A detailed description of this table for the case of NGC 5548 is provided by \cite{Mehd15}. While performing the fit we let free to vary (but tied among the pointings) the normalization of this table. 
From the fit, a flux of (1.2$\pm$0.1)$\times$10$^{-11}$ erg s$^{-1}$ cm$^{-2}$ was found for this component.
Finally, we include the effect of the Galactic extinction using the \textit{redden} model in \textit{Xspec}. The reddening is kept fixed while fitting the data to the value E(B-V)=0.021 \cite[][]{Schl11}.\\
\indent\textbf{Soft X-rays}:
The \textit{nthcomp} model \cite[][]{Zdzi96,Zyck99} is used to reproduce the soft-excess. This model provides a thermally Comptonized continuum and the high energy roll-over is parametrized by the electron temperature. In the fitting procedure the normalization, the photon index, and the warm electron temperature kT$_{\rm{wc}}$ are free to vary between the observations. In \textit{nthcomp}, we assumed the seed photons to arise from a disc-like blackbody. As a first step, the seed photon temperature (kT$_{\rm{bb}}$) was fitted for each observation. However, since the kT$_{\rm{bb}}$ were consistent with being constant, we fit the kT$_{\rm{bb}}$ temperature tying its value among the various pointings. In P1 two ionized warm absorbers and a photoionised emission component were confirmed to contribute to the NGC4593 soft X-ray emission, thus, in the present modelling, we account for these spectral components adopting tables in \textit{Xspec} (\textit{mtable(WA1)}, \textit{mtable(WA2)} and \textit{atable(REFL)}). These tables are computed thanks to the spectral synthesis code \textit{CLOUDY}  \cite[v13.03][]{Ferl13}. The best-fit values for these components are shown in P1 (its Sect. 3.1, Tab.~3). In this paper, we adopt the same best-fit values published in P1 keeping them frozen during the fit.
Moreover, it is well known that in the soft band some spectral features cannot be directly attributed to the targeted source. For instance, spectral features close to the Si K-edge (E=1.84 keV) and the Au M-edge (E$\sim 2.4$ keV) may be an artefact of the detector systematic calibration uncertainties. To avoid these issues, we ignored the spectral bins in the energy range 1.8-2.4 keV \cite[see e.g.][]{Kaas11,DiGe15,Ursi15,Capp16,Middei18}.
However, after this procedure, a line-like feature still remains in the \textit{pn} spectra, and even if very weak (EW=8 eV), it is significant in terms of $\chi^2$, as a consequence of the high number of counts in the soft band. A single narrow Gaussian line untied and free to vary among the observations at $\sim$0.6 keV is enough to correct this residual narrow feature \cite[see also e.g.][]{Kaas11,DiGe15,Ursi15,Capp16}.
\\
\indent\textbf{Hard X-rays:} The primary continuum in the 2-79 keV band is then described by a second \textit{nthcomp} where the seed photons are assumed to arise from a disc-blackbody whose  temperature is fixed to the \textit{nthcomp} model of the Soft X-rays.
On the other hand in P1, the hard X-ray emission of NGC 4593 displays additional complexities. In fact, a prominent \ion{Fe} K$\alpha$ is found to be the superposition of a relativistically broadened and narrow components, thus we add to the primary Comptonisation continuum the reflection models \textit{relxillcp} and \textit{xillvercp} \cite[][]{Garc14a,Daus16}. These models self-consistently incorporate fluorescence emission lines and the corresponding accompanying Compton reflection hump. Then, \textit{xillvercp} supplies the narrow line component, while \textit{rellxillcp} accounts for the relativistic effects and the broad line component.
We assumed the same iron abundance A$_{\rm{Fe}}$ for \textit{xillvercp} and \textit{relxillcp}, and, in the fitting procedure, we let this parameter free to vary but tied between the various pointings. We tied the photon index and the electron temperature of the hot corona  kT$_{\rm{hc}}$ between \textit{nthcomp} and \textit{relxillcp}, while the normalizations are free to vary and untied between models and observations.
In P1 the narrow line component coming from cold material is found to be constant among the various observations thus for \textit{xillvercp} the photon index and normalizations are free to vary but tied between the observations. Moreover, since the source is spectrally variable, there is no reason to assume that the radiation incident on the distant reflector is the same as that from the primary component. We therefore allow for different photon index between \textit{xillvercp} and \textit{relxillcp}. In P1 the inner radius is found consistent with being constant among the pointings, similar to the ionization parameter $\xi$, which best-fit values are found to be R$_{\rm{in}}\simeq$40 r$_{\rm{g}}$ and $\log\xi\simeq$3 [$\log$(erg cm s$^{-1}$)]. While fitting, we let free to vary but tied between the observations both R$_{\rm{in}}$ and $\log\xi$. We show the Comptonising and the reflection components in Fig.~\ref{best_mo}.
The primary kT$_{\rm{bb}}$ is tied to the same parameter of the \textit{nthcomp} component used for reproducing the soft-excess.\\ Then, we end up with the following model:
\newline

$
\textit{redden}\times~\textit{phabs}\times~\textit{const}\times~\textit{mtable(WA1)}\times~\textit{mtable(WA2)}\times
\times[gauss+~\textit{atable(small\_BB)}+~ \textit{nthcomp$_{\rm{wc}}$}+ 
\textit{atable(REFL)} + +\textit{xillvercp}
+~\textit{nthcomp$_{\rm{hc}}$}+~\textit{relxillcp}]
$~.
\newline

\begin{figure*}
	\includegraphics[width=0.49\textwidth]{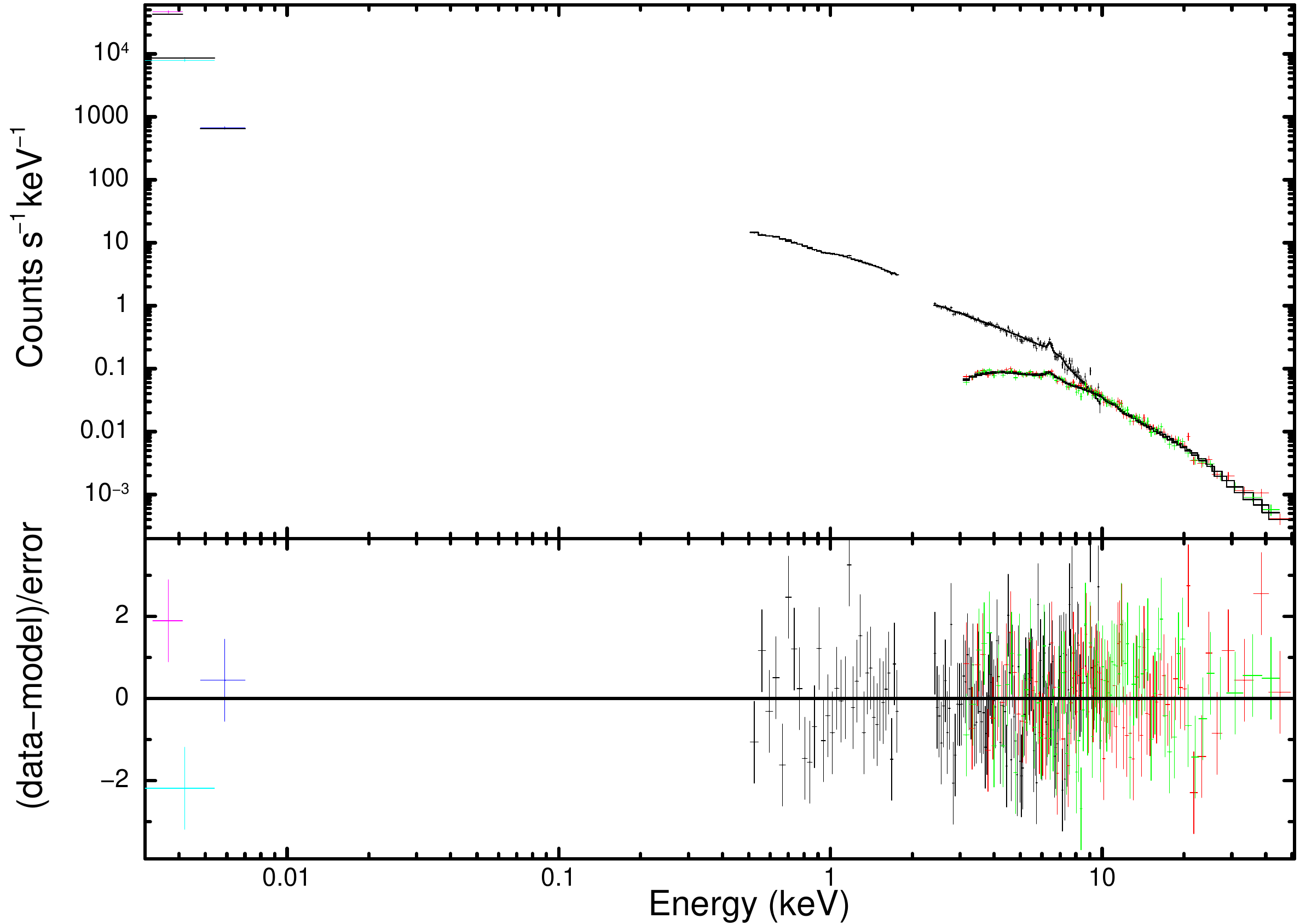}
	\includegraphics[width=0.49\textwidth]{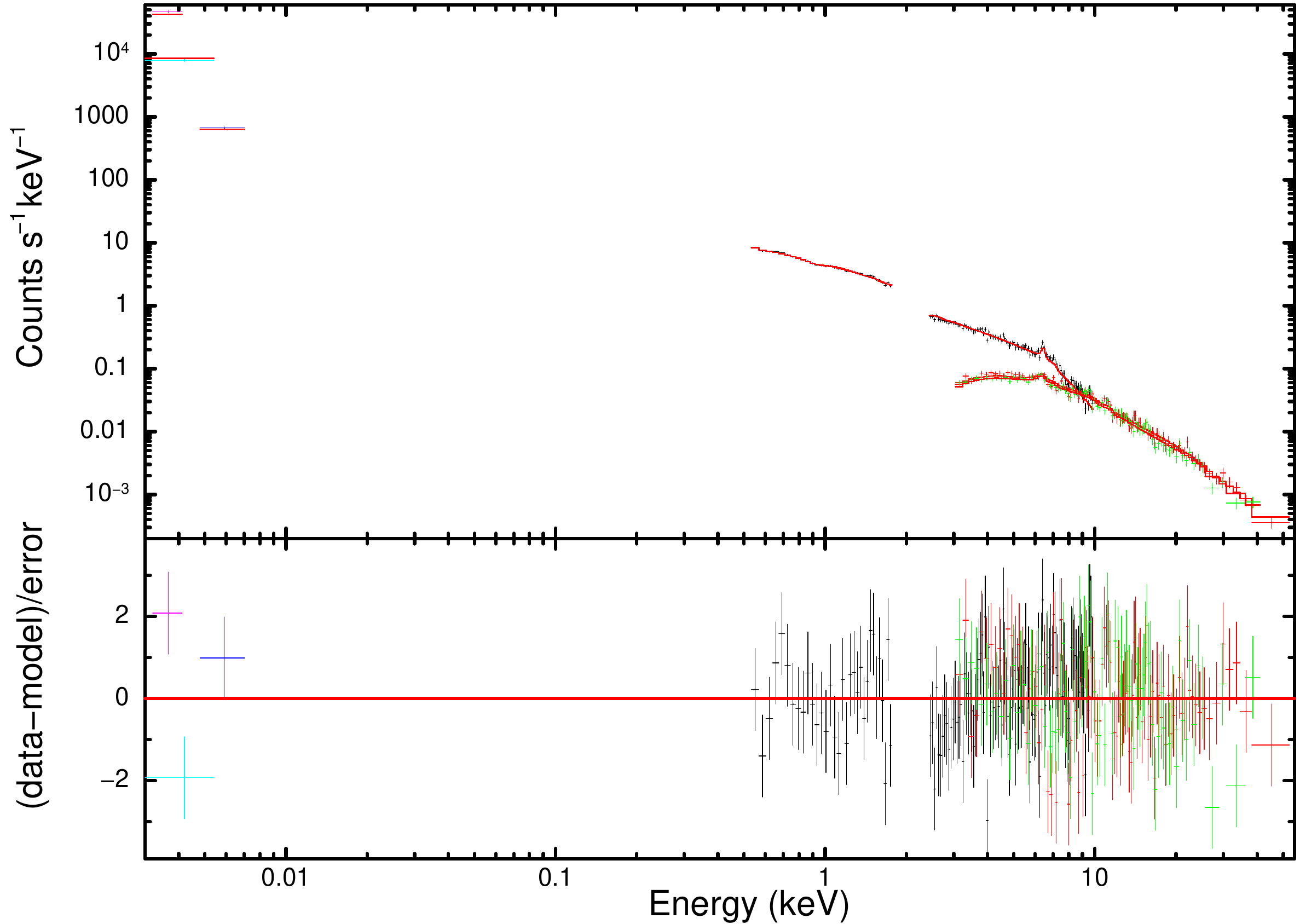}
		
	\includegraphics[width=0.49\textwidth]{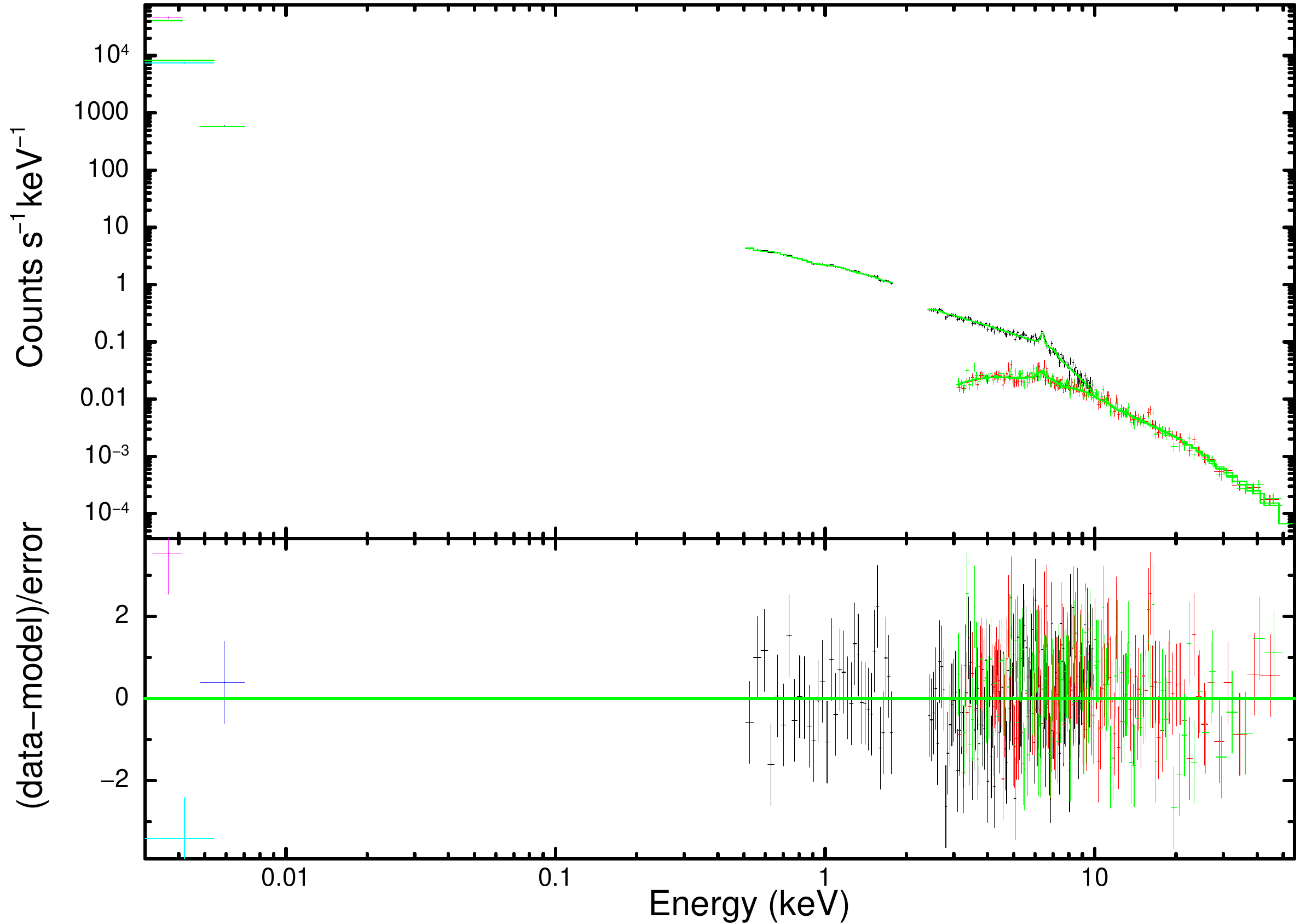}
	\includegraphics[width=0.49\textwidth]{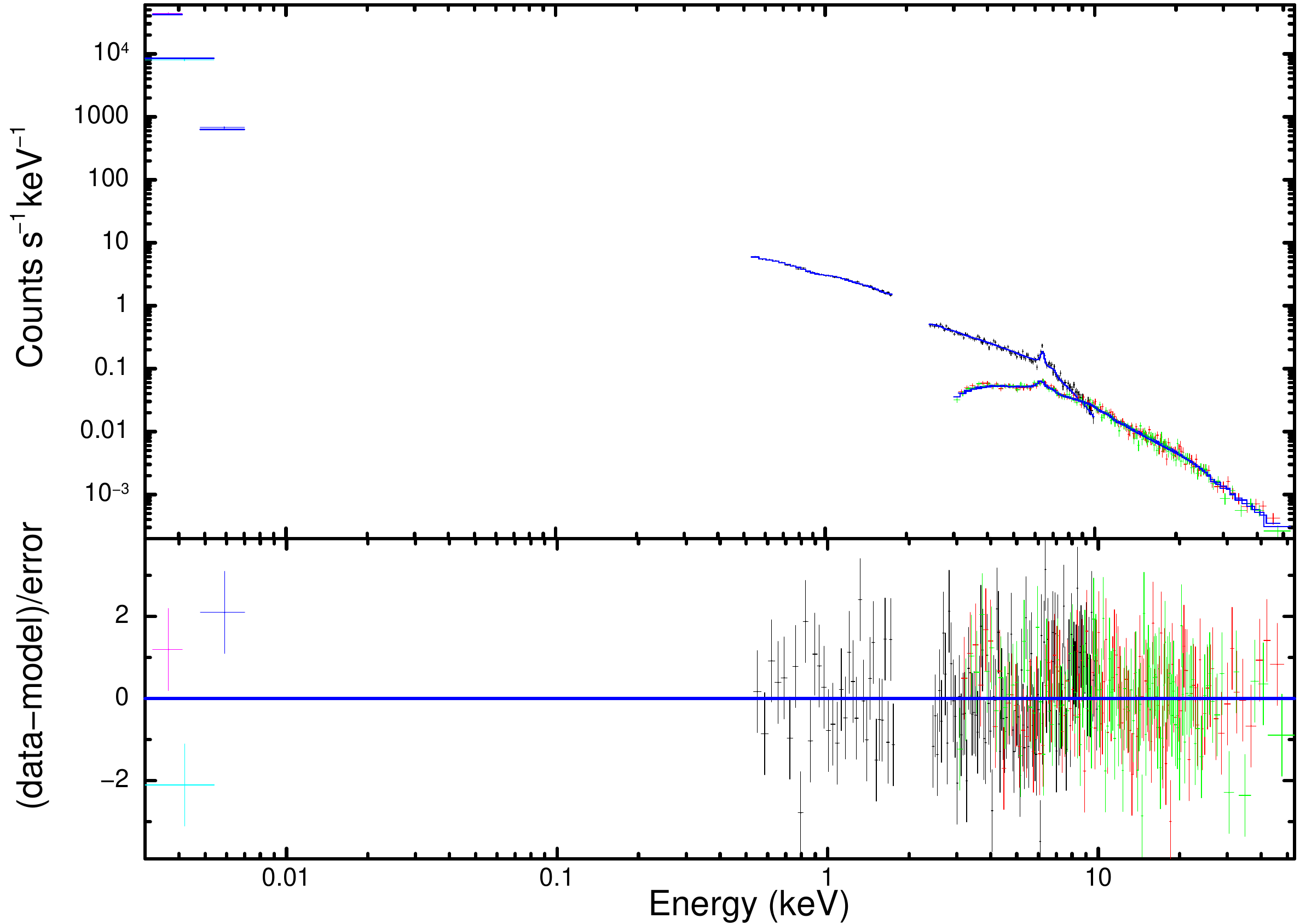}
				
    \includegraphics[width=0.49\textwidth]{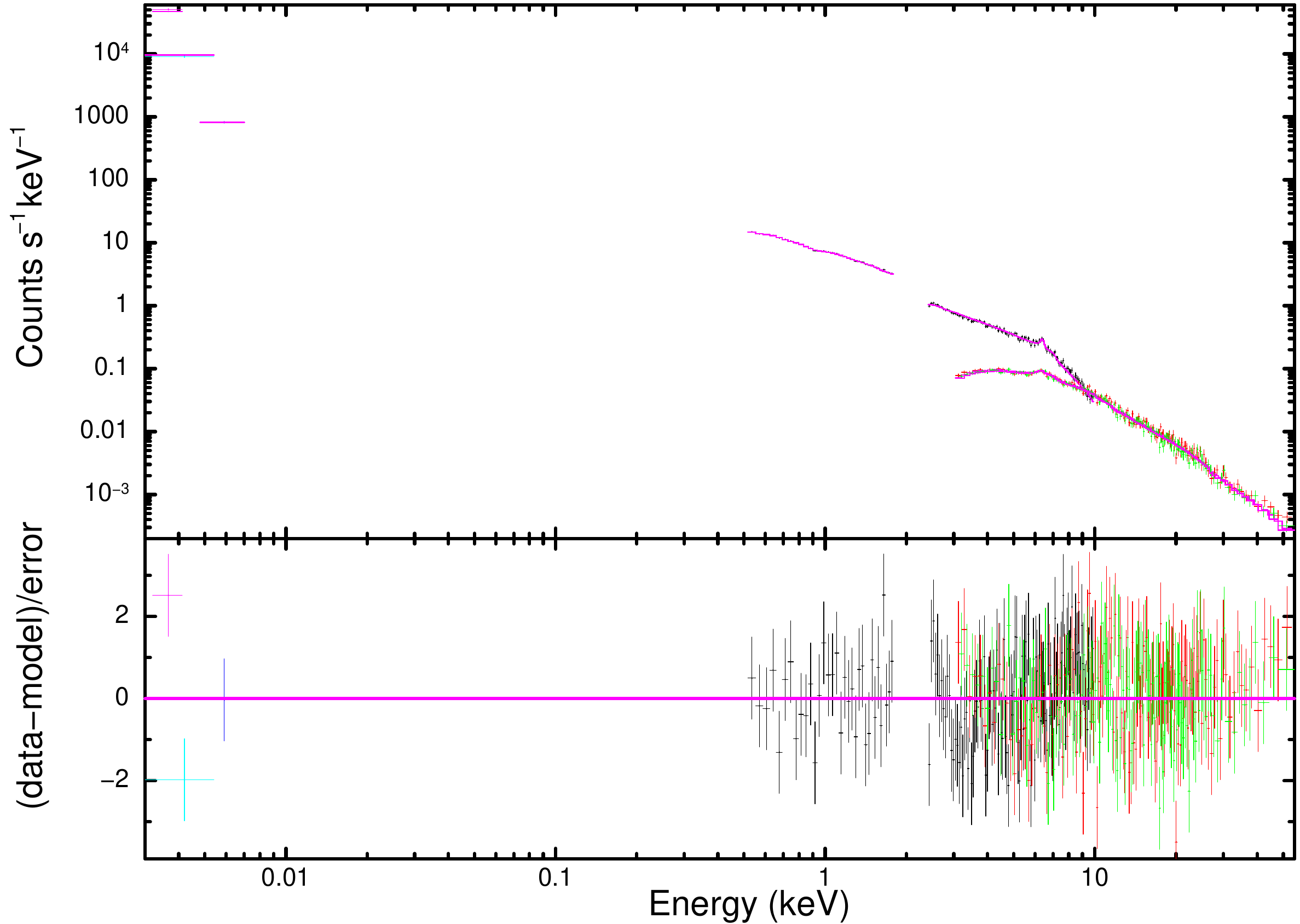}
	\includegraphics[width=0.49\textwidth]{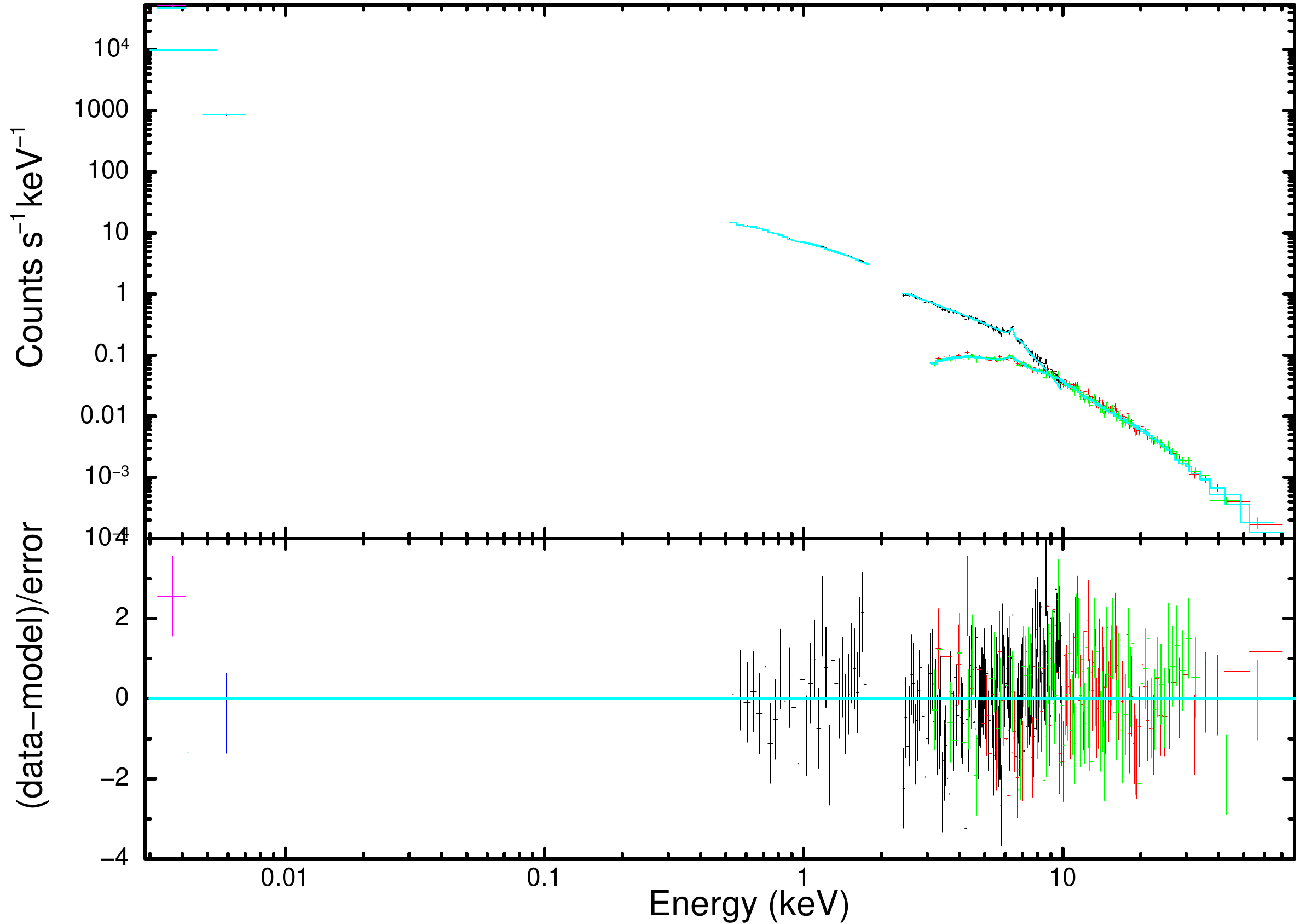}
	
	\caption{\small{The broadband best-fit corresponding to $\chi^2$=2357 for 2189 d.o.f. is showed for all the observations in the top part of the graphs while the residuals with respect to the errors are displayed in the bottom sub-panels.}\label{best}}
\end{figure*}

\begin{figure}
	\includegraphics[width=0.48\textwidth]{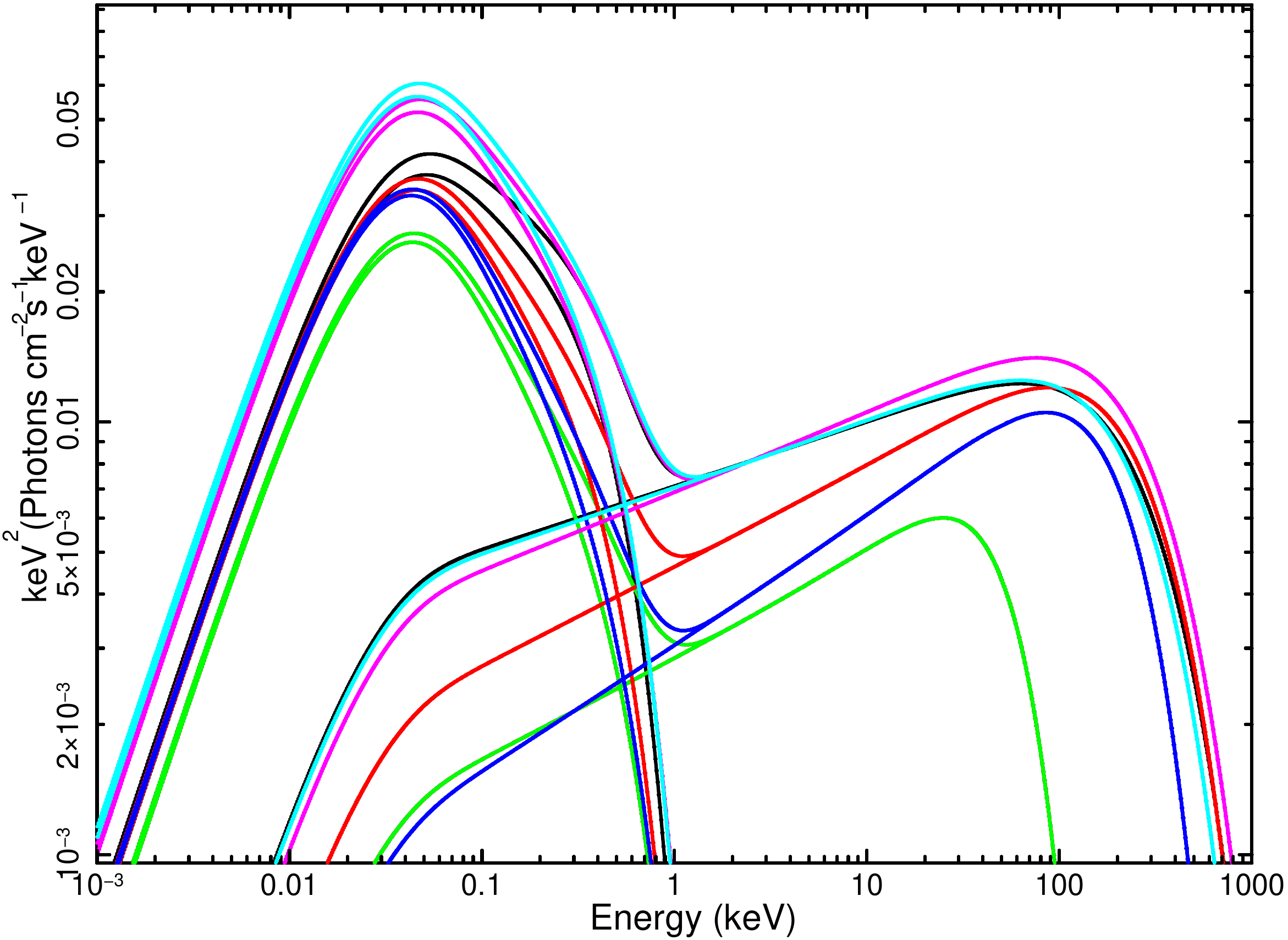}
	\caption{\small{The warm and hot Comptonising components for the various observations.\label{compt_all}}}
\end{figure}

\begin{figure}
	\includegraphics[width=0.5\textwidth]{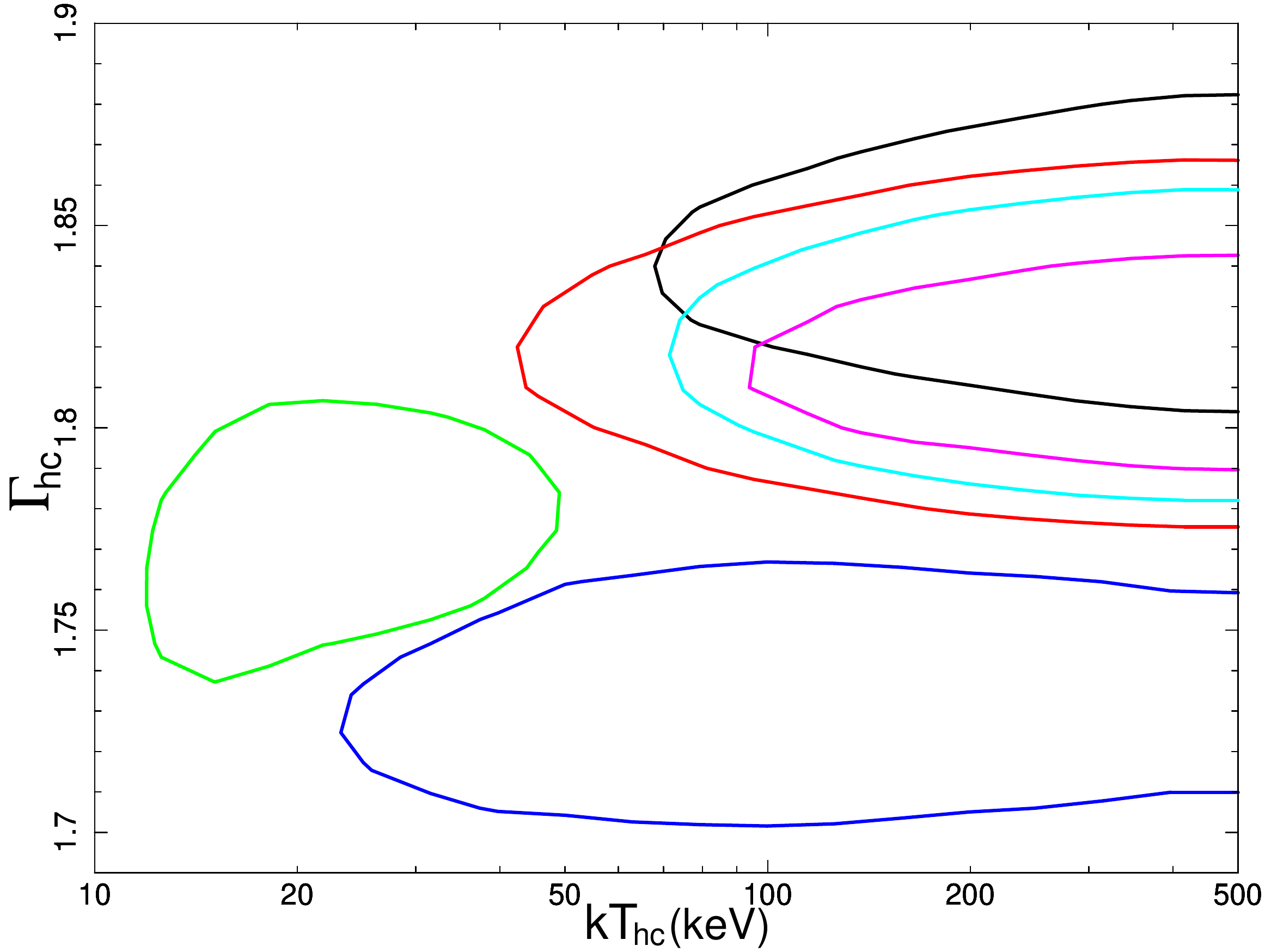}
	\includegraphics[width=0.5\textwidth]{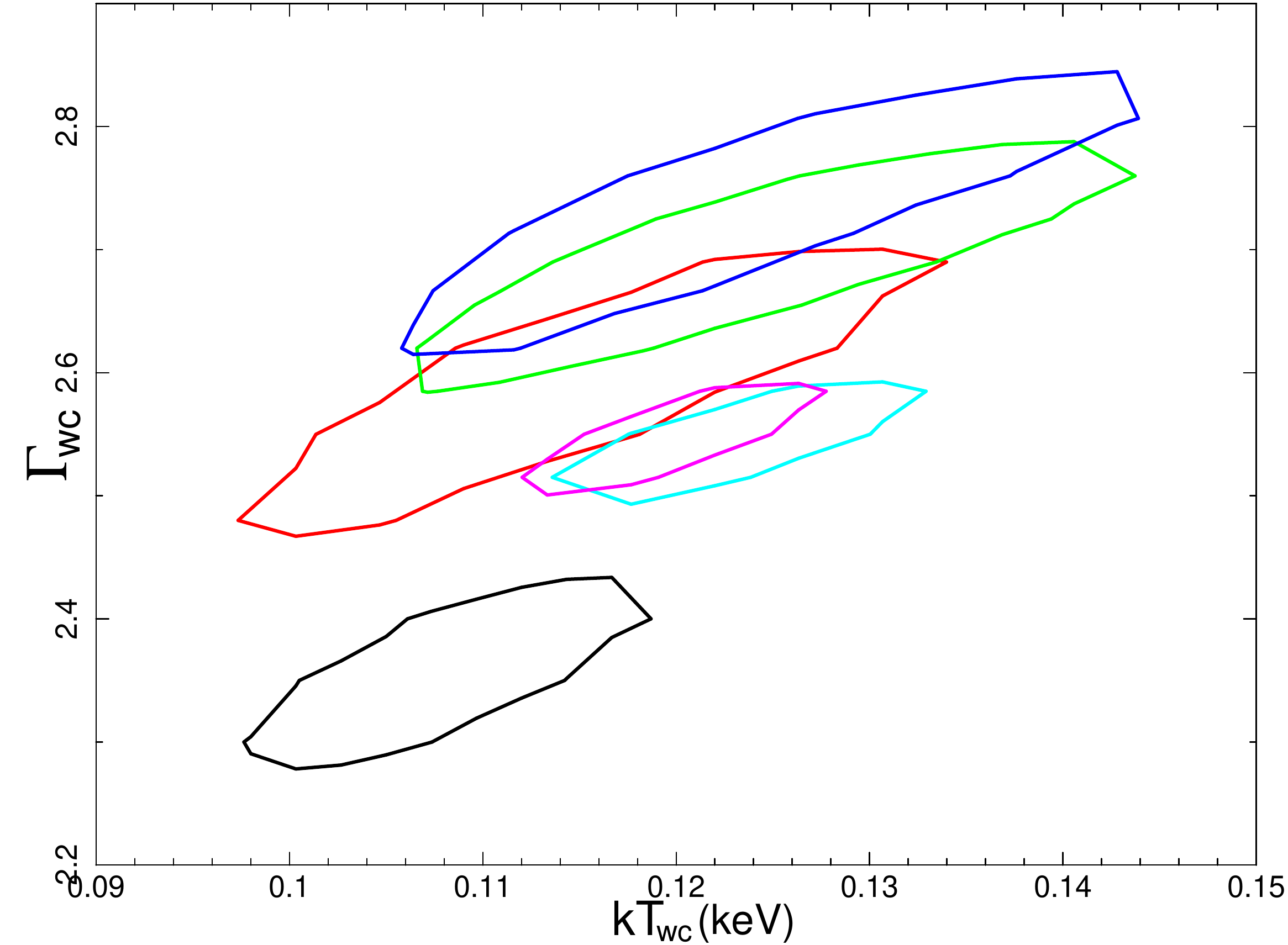}
	\caption{\small{Confidence regions at 90\% confidence level for the photon index and hot corona temperature (top panel), and for the warm corona.}}\label{cont}
\end{figure}

\begin{figure}
	\includegraphics[width=0.5\textwidth]{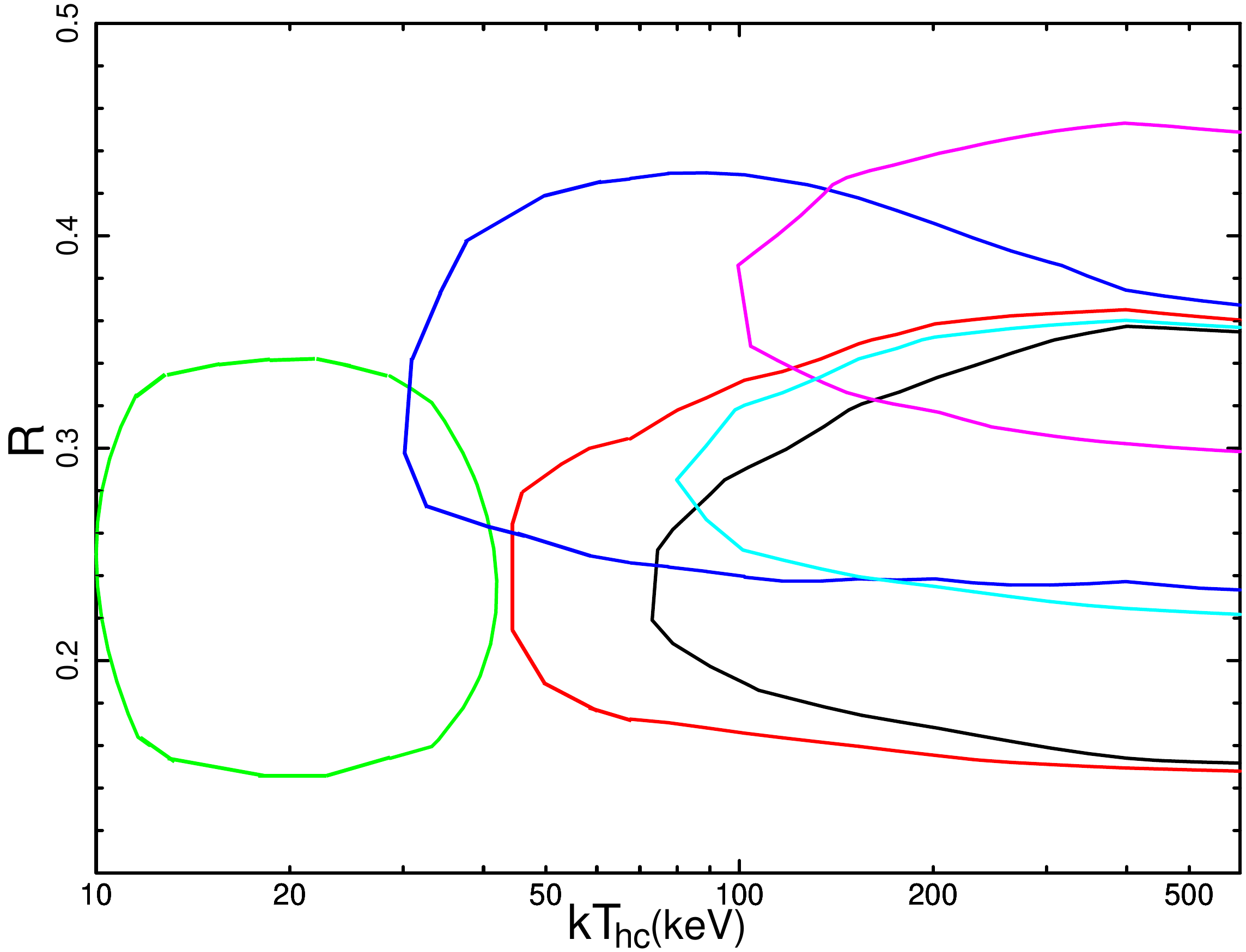}
	\caption{\small{Contours at 90\% confidence level for the reflection fraction and the hot corona temperature (top panel). Contours are flat, thus no degeneracy is present between the parameters.}}\label{refl_cont}
\end{figure}

\begin{figure}
	\includegraphics[width=0.48\textwidth]{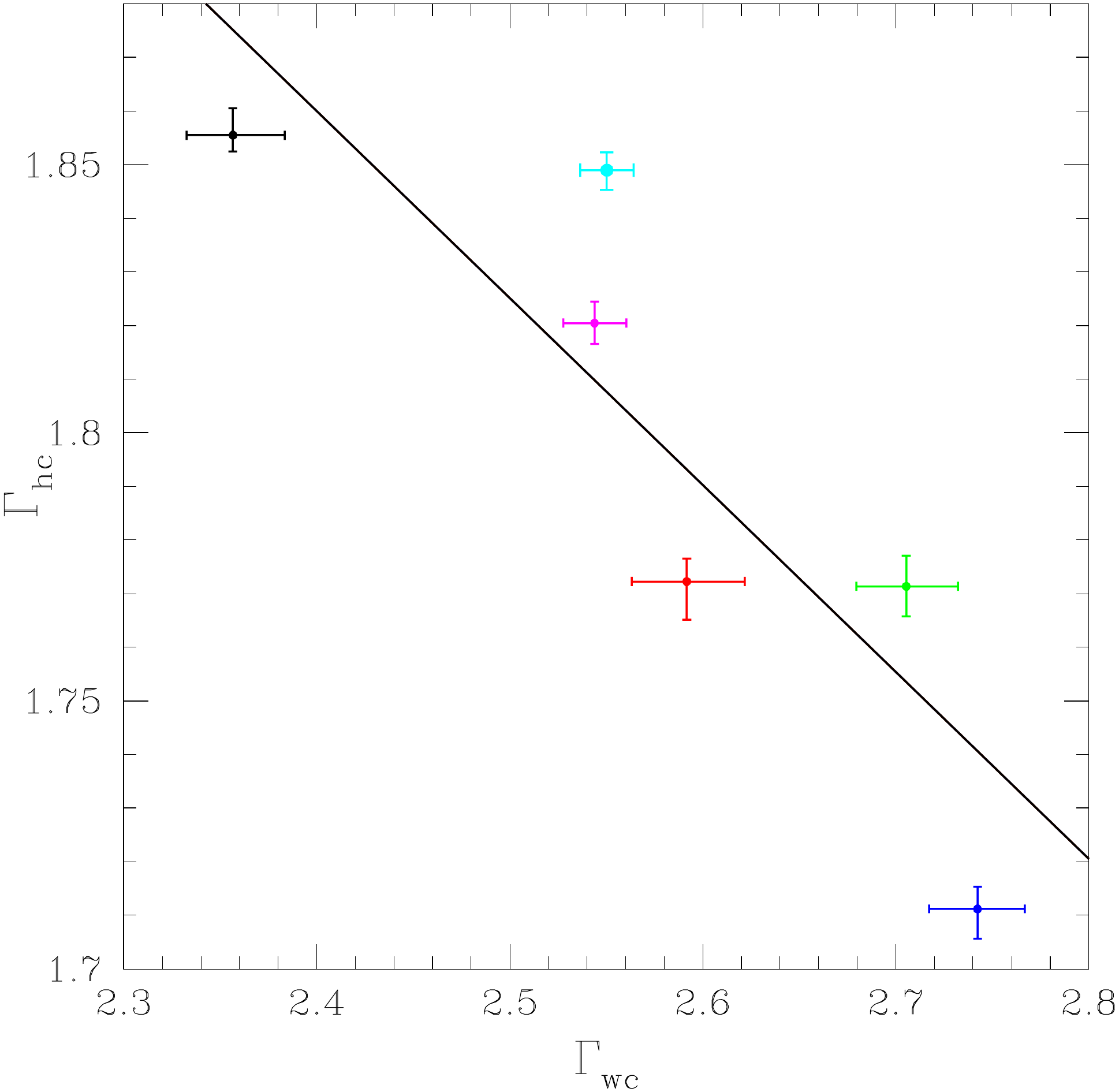}
	\includegraphics[width=0.48\textwidth]{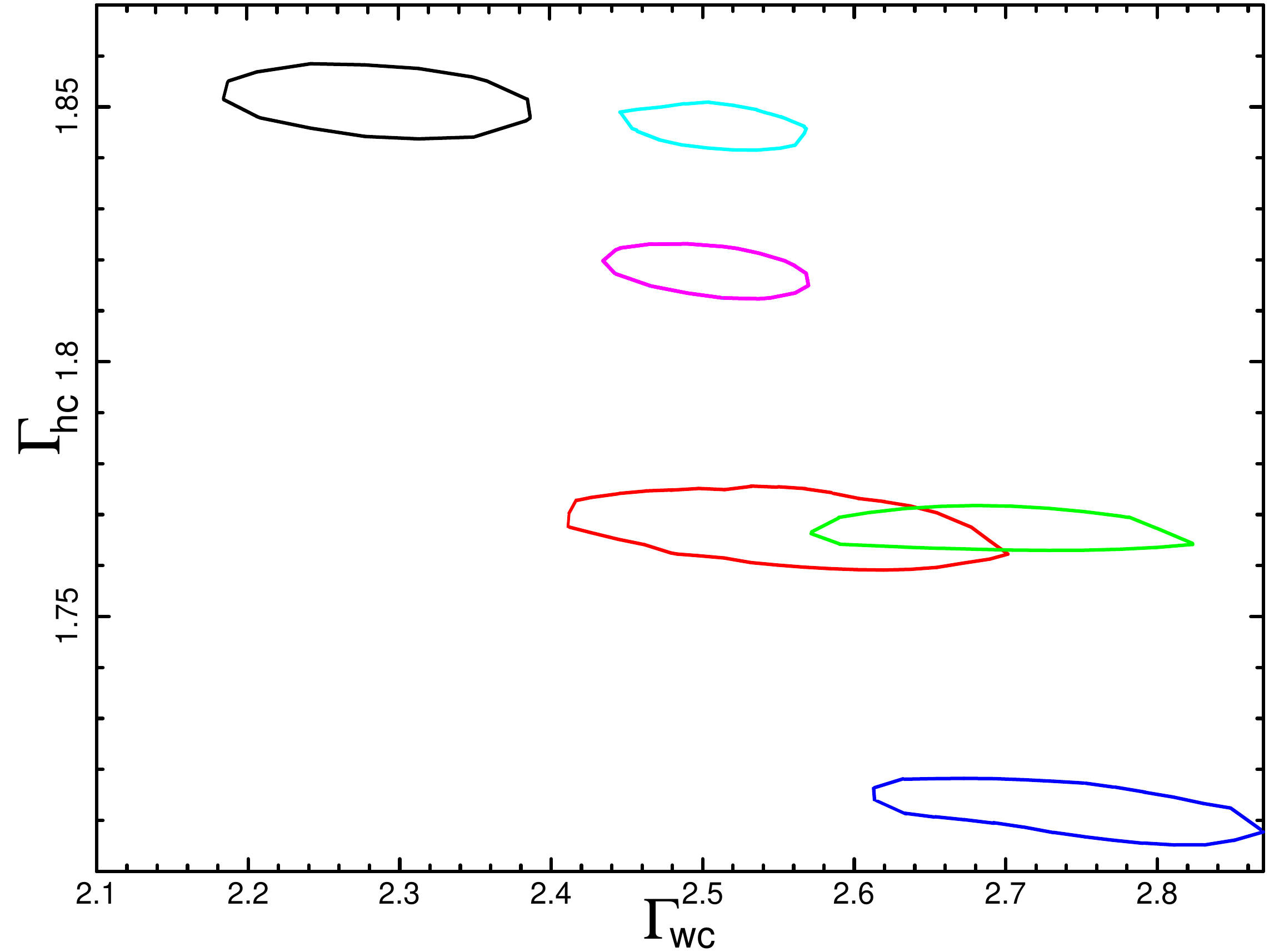}
	\caption{\small{\textit{Top panel}: The photon index for the hot corona and warm corona are reported. The solid line represent the fit to the points and the associated Pearson coefficient and null probability are also shown.
	\textit{Bottom panel}: Contours at 90\% confidence level of both $\Gamma_{\rm{hc}}$ and $\Gamma_{\rm{wc}}$ for each observation. Flat contours indicate that the two interesting quantities are measure independently, therefore the correlation is not due to the presence of degeneracy.	
    \label{gamgam}
	}}
\end{figure}

\noindent The adoption of this model results in a best-fit of $\chi^2$=2357 for 2189 d.o.f., see Fig.~\ref{best}. The values of the best-fit parameters are reported in Tab. 2. 
A super Solar iron abundance A$_{\rm{Fe}}$=2.8$^{+0.2}_{-0.3}$ is required by the fit, and this value is in agreement with the abundance quoted in P1 (A$_{\rm{Fe}}$=2.6$^{+0.2}_{-0.4}$). The disc temperature is found to be constant among the pointings with a corresponding best-fit value of kT$_{\rm{bb}}$=12$\pm$1 eV. As in P1, the hard component displays variability since $\Gamma_{\rm{hc}}$ ranges between 1.71 and 1.85, and a similar behaviour is found for the \textit{nthcomp} modelling the soft-excess ($\Delta\Gamma_{\rm{wc}}\sim$0.3), see Fig~\ref{compt_all} and Fig.~\ref{cont}. The electron temperature kT$_{\rm{hc}}$ remains unconstrained in most of the observations, apart from observation 2, where the fit returns kT$_{\rm{hc}}=17^{+20}_{-4}$ keV. However, from Fig.~\ref{cont}, hints of variability for the physical properties of the hot corona remain.
On the other hand, the warm corona temperature is consistent with being constant: kT$_{\rm{wc}}=0.12\pm0.01$ keV.
We then use the best-fit values of the hot corona and warm corona temperature to calculate the corresponding optical depths of the electron distributions. To do this, we take advantage of the \textit{nthcomp} internal routine used to compute the Thomson optical depth. This procedure assumes a spherical plasma geometry. For both the Comptonising components the best-fit values for the optical depths are reported in Tab. 2~. The optical depth of the hot component can be constrained only for observation 2,  $\tau_{\rm{hc}}$=2.1$^{+0.4}_{-0.9}$, and, for the remaining observations, upper limits are found to be in the range 0.9$\leq$$\tau_{\rm{hc}}$$\leq$3.2
The optical depth of the warm component is found to vary. The $\tau_{\rm{wc}}$ ranges between $\sim$35 and $\sim$50.\\
\indent Finally, we searched for correlations between the best-fit parameters quoted in Tab. 2.
The Pearson cross-correlation coefficients (P$_{\rm{cc}}$) and the corresponding null probability P(r>) are shown in Tab. 3. Strong anticorrelations are found between the coronal temperature and the optical depth for both the hot and warm components. A noticeable anticorrelation occur also between $\Gamma_{\rm{hc}}$ ($\Gamma_{\rm{wc}}$) and $\tau_{\rm{hc}}$ ($\tau_{\rm{wc}}$), and, interestingly, the photon index of these two components are significantly anticorrelated. In fact, as shown in Fig.~\ref{gamgam}, lower $\Gamma_{\rm{hc}}$ values correspond to steeper $\Gamma_{\rm{wc}}$. The Pearson cross-correlation coefficient is P$_{\rm{cc}}$=-0.82 with a corresponding null probability of 0.03. We have further tested this latter correlation checking for dependencies or degeneracies due to the model. Contours in Fig.~\ref{gamgam} show that the correlation is not due to model parameter degeneracy.
Moreover, we checked for the presence of additional degeneracies between the parameters, in particular focusing on the hard Comptonising component and the reflection one. However, the computed contours are flat, thus no degeneracy is present between the parameters of interest, see Fig. \ref{refl_cont}.

\begin{table*}
	\setlength{\tabcolsep}{1.pt}
	\centering 
	\caption{\small{Best-fit values for the parameters corresponding to the best-fit model ($\chi^2$=2357 for 2189 d.o.f.). This fit is also characterized by a seed photons temperature of kT$_{\rm{bb}}$=12$\pm$1 eV and an Iron abundance A$_{\rm{Fe}}$=2.8$_{-0.3}^{+0.2}$. Both these parameters were free to vary but tied between the different pointings. The luminosities and observed fluxes for the 0.3-2 keV and 2-10 keV bands are also showed.}}
	\begin{tabular}{c c c c c c c c c c c c c c}
		\hline	 
		\hline
		Obs &  $\Gamma_{\rm{hc}}$ & kT$_{\rm{hc}}$& $\tau$$_{\rm{hc}}$&N$_{\rm{hc}}$(10$^{-4}$)&N$_{\rm{rel}}$ (10$^{-5}$) &$\Gamma_{\rm{wc}}$ &kT$_{\rm{wc}}$&$\tau$$_{\rm{wc}}$ &N$_{\rm{wc}}$ (10$^{-4}$)&L$_{\rm{0.3-2}}$(10$^{42}$)&L$_{\rm{2-10}}$(10$^{42}$)&F$_{\rm{0.3-2}}$  (10$^{-10}$)&F$_{\rm{2-10}}$ (10$^{-11}$) \\
		&   & (keV)&& ph/keV/cm$^{2}$/s& ph/keV/cm$^2$/s & &keV& & ph/keV/cm$^{2}$/s&erg/s&erg/s&ergs/cm$^2$/s& ergs/cm$^2$/s \\
		\hline
		\hline
		\\
		1a  & 1.85$^{+0.01}_{-0.01}$ & >60 & <1.7 & 7.1$^{+0.2}_{-0.2}$&3.6$^{+0.6}_{-1.0}$&2.35$^{+0.04}_{-0.04}$&0.11$^{+0.01}_{-0.01}$& 47$^{+3}_{-3}$&4.9$^{+1.6}_{-0.6}$&3.68$\pm$0.02&4.61$\pm0.02$&2.05$_{-0.01}^{+0.40}$&2.56$_{-1.0}^{+0.05}$\\
		\\
		1b  & 1.77$^{+0.01}_{-0.01}$ & >80& <1.5 & 4.6$^{+0.9}_{-0.9}$&2.9$^{+0.8}_{-0.3}$&2.59$^{+0.03}_{-0.08}$&0.11$^{+0.01}_{-0.01}$&42$^{+2}_{-2}$ & 3.0$^{+1.2}_{-1.1}$&2.34$\pm$0.02 &3.47$\pm0.03$&1.29$_{-0.01}^{+0.30}$&1.92$_{-0.90}^{+0.12}$\\
		\\
		2  & 1.77$^{+0.01}_{-0.01}$ & 17$^{+20}_{-4}$ & 2.1$^{+0.4}_{-0.9}$ & 2.8$^{+0.8}_{-1.0}$&1.3$^{+0.3}_{-0.3}$&2.74$^{+0.05}_{-0.05}$&0.14$^{+0.01}_{-0.01}$&35$^{+3}_{-2}$ & 2.7$^{+0.7}_{-0.4}$&1.49$\pm$0.02&2.45$\pm0.03$&0.82$_{-0.01}^{+0.01}$&1.37$_{-0.02}^{+0.02}$\\
		\\
		3  &1.71$^{+0.01}_{-0.01}$ & >30 & <3.2& 3.0$^{+1.0}_{-0.5}$&2.9$^{+0.6}_{-0.9}$&2.74$^{+0.05}_{-0.04}$&0.12$^{+0.01}_{-0.01}$&38$^{+3}_{-2}$&2.7$^{+0.7}_{-0.6}$&1.66$\pm$0.02 &2.62$\pm0.03$&0.92$_{-0.11}^{+0.3}$&1.46$_{-0.08}^{+0.02}$\\
		\\
		4  & 1.82$^{+0.01}_{-0.01}$ &>150 & <0.9 & 6.9$^{+0.2}_{-0.2}$&5.7$^{+0.7}_{-0.8}$&2.54$^{+0.03}_{-0.01}$&0.12$^{+0.01}_{-0.01}$& 41$^{+2}_{-2}$&7.6$^{+0.1}_{-0.1}$&3.86$\pm$0.02&4.91$\pm$0.02&2.14$_{-0.01}^{+0.80}$&2.73$_{-0.06}^{+0.04}$\\
		\\
		5  & 1.85$^{+0.01}_{-0.01}$ & >140 & <0.9 & 7.0$^{+1.2}_{-1.2}$&4.2$^{+0.6}_{-0.8}$&2.55$^{+0.03}_{-0.02}$&0.12$^{+0.01}_{-0.01}$& 41$^{+2}_{-2}$&7.3$^{+0.5}_{-0.6}$&3.82 $\pm$0.02&4.67$\pm$0.02&2.12$_{-0.01}^{+0.50}$&2.60$_{-0.10}^{+0.03}$\\
		\hline
	\end{tabular}
\end{table*}

\begin{figure}
	\includegraphics[width=0.48\textwidth]{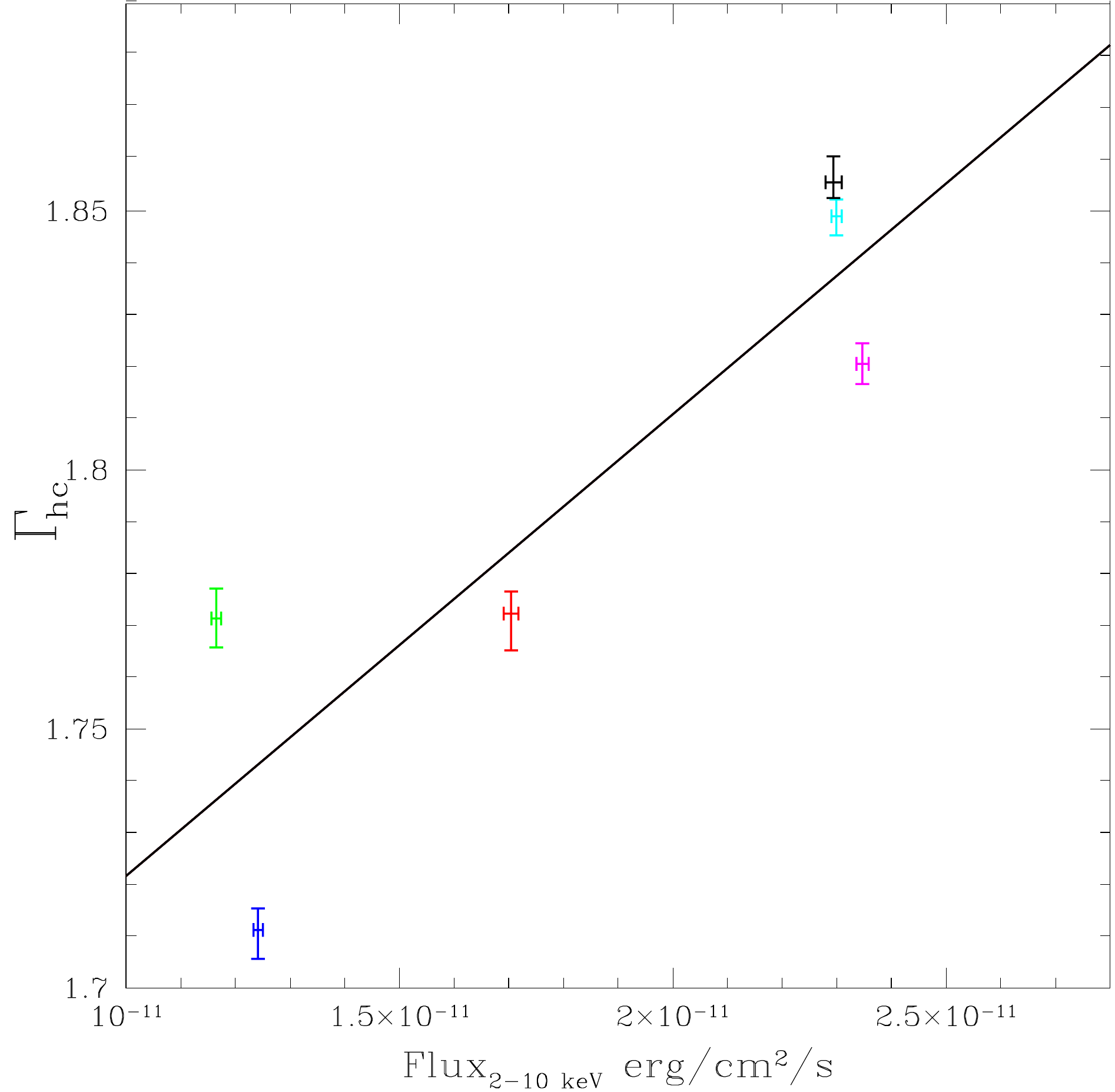}
	\includegraphics[width=0.48\textwidth]{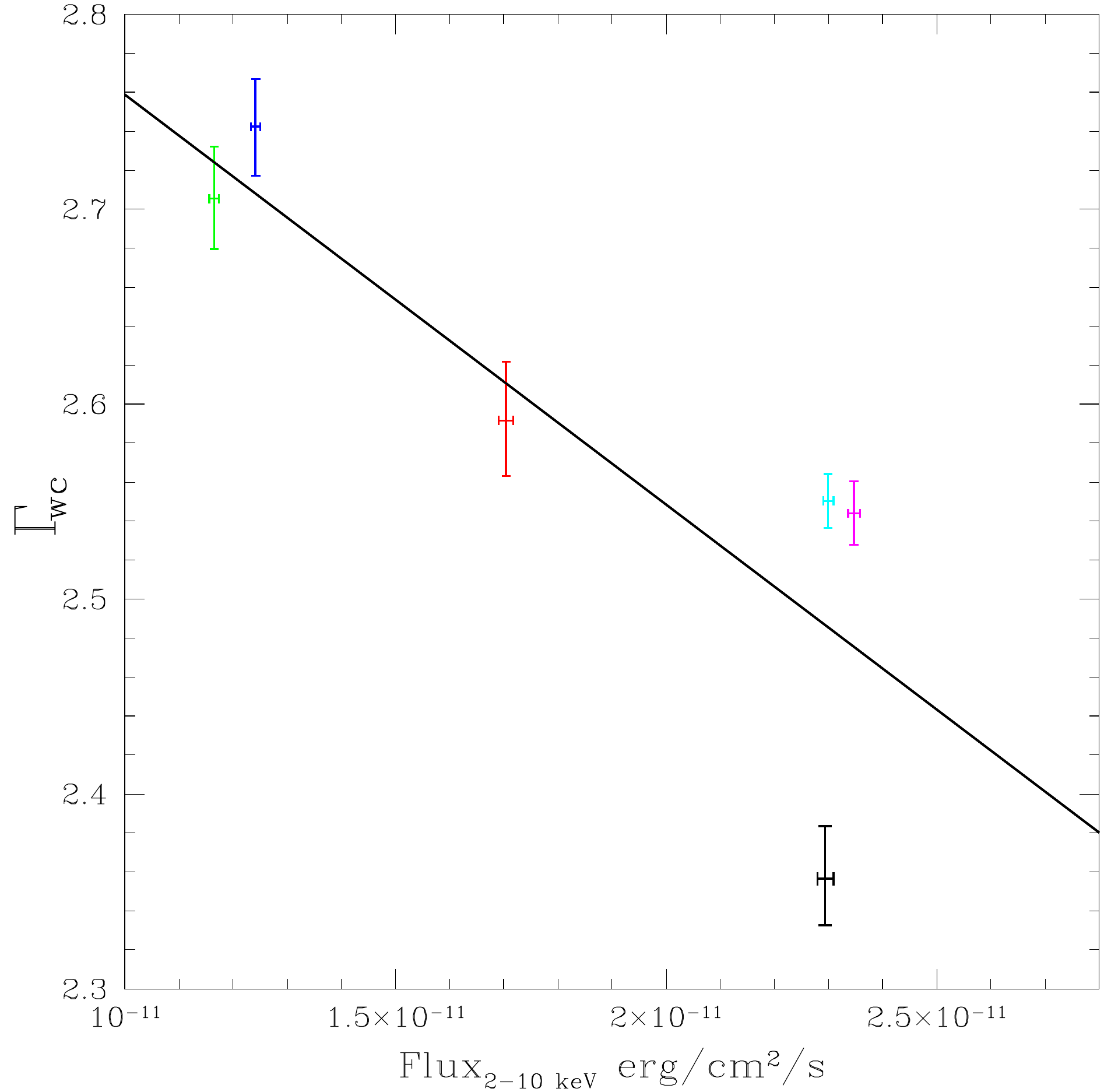}
	\caption{\small{The photon index of the hot and warm corona are shown as a function of the 2-10 keV hot component flux. The photon index for both the components are found correlated with the 2-10 keV flux with corresponding Pearson coefficients P$_{\rm{cc}}$=0.88 (P(>r)=0.03) and P$_{\rm{cc}}$=-084 (P(>r)=0.03) for the $\Gamma_{\rm{hc}}$ and $\Gamma_{\rm{wc}}$}, respectively.}\label{gamflux}
\end{figure}
\begin{table*}
	\centering 
	\setlength{\tabcolsep}{9.pt}
	\caption{\small{The Pearson cross-correlation coefficients and their corresponding null hypothesis probabilities are reported for all the best-fit values of the parameters quoted in Tab. 2. The X is used when the two parameters of interest are not significantly correlated (P$_{\rm{cc}}$<$\pm$0.70).}}
	\begin{tabular}{c c c c c c c c c c c c c c c}
		\hline
		\\
		P$_{\rm{cc}}$&$\Gamma_{\rm{hc}}$ & kT$_{\rm{hc}}$& $\tau$$_{\rm{hc}}$&N$_{\rm{hc}}$&N$_{\rm{rel}}$&$\Gamma_{\rm{wc}}$ &kT$_{\rm{wc}}$&$\tau$$_{\rm{wc}}$ &N$_{\rm{wc}}$&L$_{\rm{0.3-2}}$&L$_{\rm{2-10}}$&F$_{\rm{0.3-2}}$  &F$_{\rm{2-10}}$ \\
		P(r>)&&&&&&&&&&&&&\\
		\hline
		
		$\Gamma_{\rm{hc}}$ &-&X&-0.82&0.9&X&-0.82&X&X&0.77&0.89&0.87&0.89&0.87&\\
		&&&0.04&0.03&&0.04&&&0.05&0.03&0.03&0.03&0.03&\\
		\hline
		kT$_{\rm{hc}}$&X&-&-0.85&0.80&0.88&X&X&X&0.91&0.84&0.85&0.83&0.85&\\
		&&&0.04&0.05&0.03&&&&0.02&0.04&0.04&0.04&0.04&\\
		\hline
		$\tau$$_{\rm{hc}}$&-0.82&-0.85&-&-0.80&X&X&X&X&-0.79&-0.79&-0.81&-0.79&-0.81&\\
		&0.04&0.04&&0.05&&&&&0.05&0.05&0.05&0.05&0.05&\\
		\hline
		N$_{\rm{hc}}$&0.9&0.80&-0.80&-&0.81&-0.87&X&0.77&0.87&0.99&0.99&0.99&0.99&\\
		&0.03&0.05&0.05&&0.05&0.03&&0.06&0.03&0.01&0.01&0.01&0.01&\\
		\hline
		N$_{\rm{rel}}$&X&0.88&X&0.81&-&X&X&X&0.88&0.85&0.87&0.85&0.87&\\
		&&0.03&&0.05&&&&&0.03&0.04&0.03&0.04&0.03&\\
		\hline
		$\Gamma_{\rm{wc}}$&-0.82&X&X&-0.87&X&-&X&-0.95&X&-0.82&-0.83&-0.83&-0.83&\\
		&0.04&&&0.03&&&&0.02&&0.04&0.04&0.04&0.04&\\
		\hline
		kT$_{\rm{wc}}$&X&X&X&X&X&X&-&-0.86&X&X&X&X&X&\\
		&&&&&&&&0.04&&&&&&\\
		\hline
		$\tau_{\rm{wc}}$&X&X&X&0.77&X&-0.95&-0.86&-&X&X&X&X&X&\\
		&&&&0.06&&0.02&0.04&&&&&&&\\
		\hline
		N$_{\rm{wc}}$&0.77&0.91&-0.79&0.87&0.88&X&X&X&-&0.91&0.90&0.91&0.91&\\
		&0.05&0.02&0.05&0.03&0.03&&&&&0.02&0.03&0.02&0.03&\\
		\hline
		L$_{\rm{0.3-2}}$&0.89&0.84&-0.79&0.99&0.85&-0.82&X&X&0.91&-&0.99&0.90&0.95&\\
		&0.03&0.04&0.05&0.01&0.04&0.04&&&0.02&&0.02&0.02&0.02&\\
		\hline
		L$_{\rm{2-10}}$&0.87&0.85&-0.81&0.99&0.87&-0.83&X&X&0.90&0.99&-&0.89&0.96&\\
		&0.03&0.04&0.05&0.01&0.03&0.04&&&0.03&0.02&&0.03&0.02\\
		\hline
		F$_{\rm{0.3-2}}$&0.89&0.83&-0.79&0.99&0.85&-0.83&X&X&0.91&0.90&0.89&-&0.99&\\
		&0.03&0.04&0.05&0.01&0.04&0.04&&&0.02&0.02&0.03&&0.01&\\
		\hline
		F$_{\rm{2-10}}$&0.87&0.85&-0.81&0.99&0.87&-0.83&X&X&0.91&0.95&0.96&0.99&-&\\
		&0.03&0.04&0.05&0.01&0.03&0.04&&&0.03&0.02&0.02&0.01&&\\
		\hline
		\hline
	\end{tabular}
\end{table*}

\begin{figure}
	
	\includegraphics[width=0.48\textwidth]{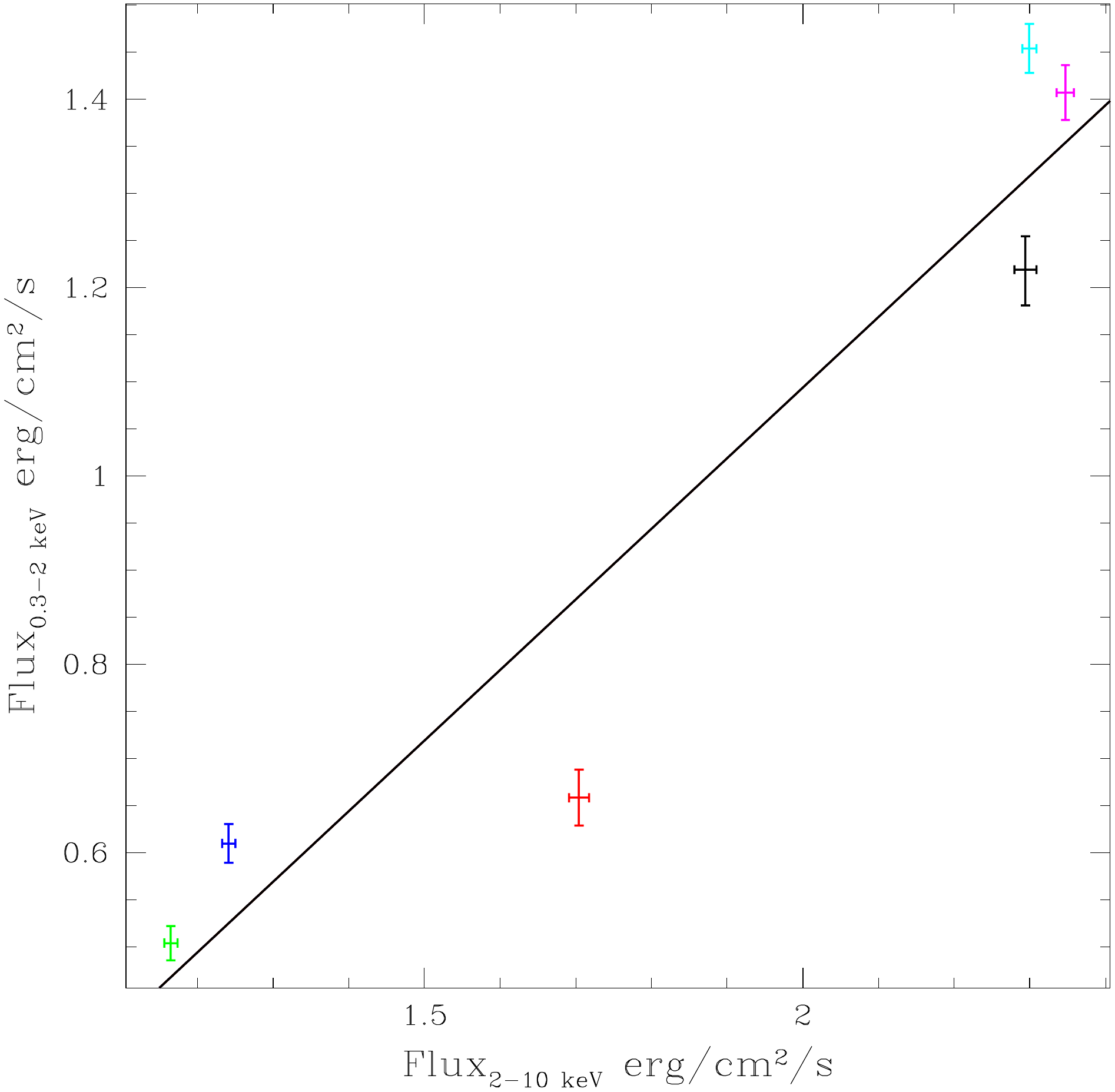}
	\caption{\small{The primary X-ray flux (2-10 keV) of the hot
			\textit{nthcomp} component  versus  the  flux  (0.3-2  keV)  of  the  warm
			\textit{nthcomp} component, as measured in the different  observations. The linear fit to the data is displayed by the solid black line (P$_{\rm{cc}}$=0.95 and P(>r)=0.02)\label{fluxflux}}. Fluxes are in units of 10$^{-11}$ erg/cm$^2$/s.}
\end{figure}

\section{Discussion}

We have performed a multiwavelength spectral investigation on the high S/N data of five simultaneous \textit{XMM-Newton} and \textit{NuSTAR} observations of NGC 4593.
As commonly observed in Seyfert galaxies \cite[e.g.][]{Pico05,Bian09,Scot12}, the broadband emission of NGC 4593 displays the presence of two main components, a primary power law and a soft-excess that becomes prominent below $\sim$ 1 keV (see P1).\\
\indent Both components show strong variability (see Fig. 1 of P1). Flux variations are typical of AGN activity, and in the X-rays, they are observed down to timescales of hours. On such short timescales, variability can be used to probe the innermost regions of AGN. Several authors showed that this variability is correlated with the BH mass \cite[e.g.][]{Nand97,czerny01,Vaug03,Niko04,McHa06}. We then follow \cite{Ponti12} to compute the normalised excess variance and its associated error for NGC 4593 obtaining $\sigma_{\rm{rms}}^2$=0.003$\pm$0.002 in the 2-10 keV band. Adopting the relation between $\sigma_{\rm{rms}}^2$ and M$_{\rm{BH}}$, we estimate the NGC 4593 BH mass to be M$_{\rm{BH}}$=(5.8$\pm$2.1)$\times$10$^{6}$ M$_\odot$. Our estimate is marginally compatible with the reverberation mapping value provided by \cite{Denn06}: M$_{\rm{BH}}$=(9.8$\pm$2.1)$\times$10$^{6}$ M$_\odot$.\\
\indent Moreover, in the present study we show that the optical-UV/X-rays emission of NGC 4593 can be explained in terms of two Comptonising components, a warm optically thick corona, and a hot optically thin medium.
The soft-component photon index shows significant variations between the different observations (2.35$\lesssim\Gamma_{\rm{wc}}\lesssim$2.74, see Fig.~\ref{compt_all}), while the corresponding electron temperature displays a more constant behaviour (on average kT$_{\rm{wc}}$=0.12$\pm$0.01 keV). Therefore, the observed variability has to depend on $\tau_{\rm{wc}}$, which we estimate to vary in the range 35<$\tau_{\rm{wc}}$<47. On the other hand, also the hot component displays remarkable spectral variations (see Fig.~\ref{compt_all}), with variability found for both the hot electron temperature and $\tau_{\rm{hc}}$.\\
For the hot corona, kT$_{\rm{hc}}$ and $\tau_{\rm{hc}}$ seem to be anticorrelated and their trend is in agreement with that reported by \cite{Tort18a}. However, our result is not very significant since it is only based on lower (upper) limits for kT$_{\rm{hc}}$ ($\tau_{\rm{hc}}$) .\\
\subsection{Warm and hot corona regions in NGC 4593}

\indent From the best fit model we can compute the total flux emitted by the warm and hot corona (F$_{\rm{tot-soft(hard)}}$), as well as the seed soft photon flux entering them and cooling them (F$_{\rm{soft(hard)}}$), see Table \ref{fcomp}.
\begin{table}
	\centering
	\setlength{\tabcolsep}{1pt}
	\caption{\small{We report in this table the best-fit fluxes entering (F$_{\rm{soft}}$ and F$_{\rm{hard}}$) and emitted (F$_{\rm{tot-soft}}$ and F$_{\rm{tot-hard}}$) by the soft and hard corona. Moreover, the flux due to relativistic reflection is also reported. All fluxes F$_{\rm{tot-soft}}$, F$_{\rm{tot-hard}}$, F$_{\rm{soft}}$, F$_{\rm{hard}}$ and F$_{\rm{rel-refl}}$ are in units of 10$^{-11}$ erg/cm$^{2}$/s, and are computed in the 0.001-1000 keV energy interval.}}\label{fcomp}
	\begin{tabular}{c c c c c c c}
		\hline
		\\
		&Obs. 1a& Obs. 1b& Obs 2. &Obs. 3& Obs. 4& Obs. 5\\
		\hline
		\hline
		F$_{\rm{soft}}$&9.2&8.9&6.9&8.9&13.4&14.6\\
		F$_{\rm{hard}}$&1.0& 0.5& 0.3& 0.3& 0.9& 1.0\\
		F$_{\rm{rel-refl}}$&1.8& 1.3& 0.8& 1.3& 2.7 & 2.1\\
		F$_{\rm{tot-soft}}$&17.8& 15.2& 11.3 &14.3 &23.6&25.6\\
		F$_{\rm{tot-hard}}$&26.0 &20.6& 9.0& 15.1& 27.2& 25.5\\

	\end{tabular}
\end{table}
\begin{table}
	\centering
	\setlength{\tabcolsep}{5.pt}
	\caption{\small{The Pearson cross-correlation coefficients and their associated null probabilities are displayed for the warm and hot corona components flux. The subscript \textit{tot} indicates the total flux emerging from the specified component, otherwise fluxes refer to the seed photons flux entering in that highlighted component. The photon index for the warm and the hot corona are also reported. When X is used, no correlation holds between the two interesting parameters.}\label{corrf}}
	\begin{tabular}{c c c c c c c c}
		\hline
		\\
		P$_{\rm{cc}}$&F$_{\rm{soft}}$           &F$_{\rm{hard}}$    &      F$_{\rm{rel-refl}}$&    F$_{\rm{tot-soft}}$&    F$_{\rm{tot-hard}}$&   $\Gamma_{\rm{wc}}$&  $\Gamma_{\rm{hc}}$ \\
		P(r>)&&&&&&&\\
		\hline
		F$_{\rm{soft}}$            &-            &   0.72   &    0.86 &    0.98 &    0.76& X&X\\
		&&0.07&0.03&0.02&0.06&&\\
		\hline
		F$_{\rm{hard}}$           &0.72      &      -         &   0.78&      0.83& 0.91&  -0.88  &  0.94\\
		&0.07&&0.05&0.04&0.02&0.03&0.02\\
		\hline
		F$_{\rm{rel-refl}}$        &    0.86&     0.79    &       -&             0.89&     0.89   &X   &X\\
		&0.03&0.04&&0.03&0.03&&\\
		\hline
		F$_{\rm{tot-soft}}$    &0.98   & 0.83    &  0.89   &-& 0.77           &          X&  X\\ 
		&0.02 &0.04&0.03&&0.06&&\\
		\hline
		F$_{\rm{tot-hard}}$     &0.76    &  0.91    &   0.89  & 0.77      &-  &-0.83  & 0.74\\
		&0.06&0.02&0.03&0.06&&0.04&0.06\\
		\hline
		$\Gamma_{\rm{wc}}$       &X    &-0.88     &X  & X&-0.83     &  -            &-0.82\\
		&&0.03&&&0.04&&0.04\\
		\hline
		$\Gamma_{\rm{hc}}$       &X   & 0.94       &X &  X&0.74   & -0.82     &-\\
		&&0.02&&&0.06&0.04&\\
		\hline
		\hline
	\end{tabular}
\end{table}
With the estimate of the coronal best-fit optical depth we can deduce the so-called Compton amplification factor $A$ \cite[see][ for details]{pop18}. This factor accounts for the ratio between the total power emitted by the warm corona and the seed soft luminosity from the accretion disc, and it can be used to estimate the geometrical properties of the Comptonising medium. In particular, following the steps described by \cite{pop18} we find that the amplification factor for the warm corona $A_{\rm{wc}}$ ranges between 1.6 and 2. An amplification $A_{\rm{wc}}\simeq$2 is theoretically
expected for an optically thick and slab corona corona fully covering a passive disc \cite[e.g.][]{Petr13}, thus this is in agreement with a scenario in which the soft-excess arises from a warm and optically thick medium being the upper layer of a nearly passive accretion disc \cite[][]{Roza15}. Again, following the procedure in \cite{pop18}, we also estimate the amplification factor corresponding to the hot component, $A_{\rm{hc}}$. In this case, the electron temperature is poorly constrained, and we estimate lower limits for this parameter. These lower limits translate to upper limits for the amplification factor of the hard corona $A_{\rm{hc}}$ found to be in the range 50-100. Following Eq. 18 and 23 in \cite{pop18}, these translate to lower limits of the patchiness factor of the hard corona which is of the order of 2/$A_{\rm{hc}}$ for an optically thin corona above a passive disc (i.e. with no intrinsic emission \cite{Petr13}). We obtain $g\textgreater$0.02~. These values are in agreement with a scenario in which an extended Thomson thick medium is responsible for the optical-UV/soft X-ray of NGC 4593, while a compact or patchy corona intercepting and Comptonising only a fraction (few \%) of the disc seed photons explains the X-ray primary continuum.\\
\indent Concerning the interplay between the warm and hot coronal emission, in Fig.~\ref{fluxflux} we show that the primary X-ray flux is tightly correlated with the flux of the soft-excess component. 
The Pearson correlation coefficient of P$_{\rm{cc}}$=0.95 (P(>r)=0.02) suggests that the soft-excess increases for an increasing primary flux.
\indent The tight correlation we find between the photon index of the hot and warm corona is quite unusual and difficult to explain. Concerning the evolution of the hot corona properties ($\Gamma_{hc}$ and flux), it is interesting to note that the softer and more luminous spectra roughly correspond to the observations with the highest UV flux. This is not a strong effect given the small variability of the UV flux, but it suggests a simple interpretation. Qualitatively, the spectral and luminosity change can be explained assuming that the outer disc/warm corona structure pushes a bit closer to the black hole (the transition radius $R_{tr}$ between the hot and the warm corona decreases), decreasing the portion of the hot corona which "sees" an increase of the UV-EUV soft radiation. This would cause a more effective cooling of the hot gas explaining the softening of the spectrum. If this interpretation is correct however, we would expect then a decrease in the hot corona temperature. This is in contradiction with the lowest value of the temperature, observed during observation 2, while the UV flux is also the lowest. Moreover, it is not easy to understand why, in these conditions, the warm corona photon index would evolve in an opposite way.\\
\indent Actually the anticorrelation of $\Gamma_{\rm{hc}}$ and $\Gamma_{\rm{wc}}$ could be a by product of the radiative coupling between the two coronae. 
We have reported  in Table \ref{corrf} the correlation coefficient between the total corona flux, the seed soft photon flux and the photon index of both coronae.
Interestingly, F$_{\rm{tot-soft}}$ strongly correlates with F$_{\rm{hard}}$ (P$_{\rm{cc}}$=0.83) . This correlation agrees with the warm corona acting as the seed photon source for the hot corona. The larger F$_{\rm{tot-soft}}$ the larger F$_{\rm{hard}}$ but, at the same time, the steeper the hot corona spectrum due to the increase of the cooling ($\Gamma_{\rm{hc}}$ increases as indeed observed).
On the other hand, $\Gamma_{\rm{wc}}$ anticorrelates quite strongly with F$_{\rm{hard}}$ (the correlation is even stronger than between $\Gamma_{\rm{wc}}$ and $\Gamma_{\rm{hc}}$). Possibly there is some feedback then in the sense that when F$_{\rm{tot-soft}}$ increases also F$_{\rm{tot-hard}}$ increases. Then, the heat deposit in the warm corona by e.g. illumination increases (indeed, F$_{\rm{tot-hard}}$ correlates with F$_{\rm{rel-refl}}$),  producing a hardening ($\Gamma_{\rm{wc}}$ decreases) of the warm corona spectrum. \\
\indent The main problem with this interpretation is the evolution of the coronal temperature. We would expect the temperature to decrease/increase when the spectrum steepens/hardens. The present observations do not show clear evolution of these temperatures. This suggests that something else may have to change, e.g. the optical depth, to keep the temperature roughly constant. \\
\indent Note that we have assumed a constancy of the warm corona geometry. Its variation is however plausible and should add another free parameter to explain the observations. Indeed, variability of the optical-UV-to-X-rays emission of NGC 4593 could result from geometrical variations of the `two coronae' but also of the `warm corona'  and outer part of the disc, not covered by the warm corona, and, potentially, also contributing  to the optical-UV emission.
A detailed analysis of the parameter space, also adopting more sophisticated models \cite[e.g. \textit{agnsed}][]{Kubota18} self-consistently accounting for the disc contribution in the frame-work of two Comptonising coronae, is however out of the scope of the present paper and is deferred to a future work.\\

\subsection{The two-corona model in the best monitored Seyfert galaxies}
\indent Simultaneous observations including optical-UV and X-ray information have been crucial in testing the two-corona model on AGN. \textit{XMM-Newton}-\textit{NuSTAR} monitorings have been performed on a handful of local Seyfert galaxies, for instance NGC 7469 \cite[][]{Bear17} and 3C 382 \cite[][]{Ursi18}. The two-corona model was already tested on these AGN \cite[][]{Middei18,Ursi18}, and they were objects of multi epoch campaigns analogously to NGC 4593.
NGC 7469 is a type 1 Seyfert galaxy characterized by a SMBH mass of (1.1$\pm$0.1)$\times$10$^7$ $M_{\sun}$ \cite[][]{Pete14}, while a BH mass of (1.0$\pm0.3$)$\times$10$^{9}$ $M_{\sun}$ is found for the broad-line radio galaxy 3C 382 \cite[][]{Faus17}, about 2 orders of magnitude larger than NGC 4593. The comparison of the present results with those of NGC 7469 and 3C 382  emphasizes the NGC 4593 peculiar behaviour.
First, phenomenologically, the emission of NGC 4593 is characterized by strong variability, both in flux and spectral shape, and its high energy cut-off is found to vary (also see P1). These behaviours were not observed, at least during the monitorings, in NGC 7469 or 3C 382. In fact, the hard X-ray emission of NGC 7469 was consistent with a constant spectral photon index ($\Gamma^{\rm{NGC~7469}}_{\rm{hc}}$=1.78$\pm$0.02) between the different pointings, and a non variable high energy cut-off was measured to be $E^{\rm{NGC~7469}}_{\rm{cut}}$=170$^{+60}_{-40}$, \cite[][]{Middei18}. In a similar fashion, 3C 382 also can be described in terms of a constant photon index ($\Gamma^{\rm{3C~382}}_{\rm{hc}}$=1.78$\pm$0.01).
Furthermore, the physical properties of the hot Comptonising component acting in NGC 4593 are found to vary (kT$_{\rm{hc}}$ and $\tau_{\rm{hc}}$), while these quantities in 
both NGC 7469 and 3C 382 displayed a more constant behaviour with  corresponding electron temperature of 
kT$^{\rm{NGC~7469}}_{\rm{hc}}$=$45^{+15}_{-12}$ keV and kT$^{\rm{3C~382}}_{\rm{hc}}\gtrsim$20 keV, respectively.\\
\indent On the other hand, the properties of the soft-excess are more similar among these AGN. In fact, NGC 4593, 3C 382 and NGC 7469 all display a soft-excess that cannot be explained in terms of ionized reflection alone \cite[P1,~][]{Ursi18,Middei18}, while a warm Comptonisation scenario characterized by a variable photon index and a constant electron temperature is favoured. Indeed, the kT$_{\rm{wc}}$ of NGC 4593, NGC 7469 and 3C 382 ($\sim$0.12 keV, $\sim$0.7 keV and $\sim$0.6 keV, respectively) are consistent with being constant during the monitorings. Remarkable $\Gamma_{\rm{wc}}$ variability is found in NGC 4593, noticeable variations are observed in the warm photon index of 3C 382 and weaker variability in the $\Gamma_{\rm{wc}}$ of NGC 7469 are measured. Moreover, the anticorrelation between the photon index of the two Comptonising component is not found in  NGC 7469 and 3C 382, making NGC 4593 peculiar in this respect.
\\
\section{Summary}
The present paper is the second article reporting on the observational campaign targeting NGC 4593. In particular we tested the two-corona model \cite[e.g.][]{Petr13,Roza15} on this rich high S/N data-set. The obtained overall scenario is consistent with the two-corona model in which two Comptonisation processes dominate the soft and hard X-ray emission of the source. We report our findings and conclusions in the following:

\begin{itemize}
	\item Strong flux variability is observed during the monitoring and variations are observed from daily down to hourly timescales. We adopted the normalised excess variance to quantify the source variability finding $\sigma_{\rm{rms}}^2$=0.003$\pm$0.002 in the 2-10 keV band. Following \cite{Ponti12}, we convert this value into an estimate of the BH mass hosted by NGC 4593, obtaining M$_{\rm{BH}}$=(5.8$\pm$2.1)$\times$10$^{6}$ M$_\odot$. This values is marginally in agreement with the reverberation mapping measure by \cite{Denn06}.

    \item A hot Comptonisation component describes the high energy spectra of our campaign. This hot corona is phenomenologically described by a variable photon index (1.71$\leq\Gamma_{\rm{hc}}\leq$1.85).
    The hot corona temperature remains unconstrained in five over six observations. On the other hand, hints of variability are found for this parameter and this is in agreement with the high-energy cut-off variability reported in P1. In turn, we find upper limits for the optical depth of the hot corona. We notice that $kT_{\rm{hc}}$ and $\tau_{\rm{hc}}$ anticorrelate (P$_{\rm{cc}}$=-0.85 P(r>)=4\%).

    \item All spectra display a remarkable soft-excess, and a warm Comptonisation model best describes this component. The warm medium is characterized by a variable photon index $2.35<\Gamma_{\rm{wc}}<2.74$ and a constant $kT_{\rm{wc}}=0.12\pm0.01$ keV. The optical depth of the warm corona is variable $35\leq\tau_{\rm{wc}}\leq47$.
    According to this analysis, most of the accretion power is released in the warm-corona than in the accretion disc.
    \item For the first time, we observe an anticorrelation between the photon indexes of the hot and the warm corona. The origin of this trend cannot be ascribed to a model degeneracy (see Fig.~\ref{gamgam}). The interpretation of such anticorrelation is not straightforward but can result from the radiative feedback between the two coronae. 
    \item The present test on the two-corona model indicates that it reliably allows for reproducing the AGN broadband emission. Indeed, besides observational differences with other sources that can be explained in terms of different physical properties (e.g. Eddington ratio), notably the two-corona model provides good representations of the data suggesting that it accounts for a common Comptonising mechanism occurring in AGN. On the other hand, the existence of a warm corona at such temperature and optical depth above the accretion disk is expected to produce emission/absorption features from the ionized matter. The fact that we do not see any of them would imply specific physical properties (e.g. turbulence) that have to be tested with accurate radiative transfer codes. 
\end{itemize}

\section*{Acknowledgements}
We  thank the referee Aya Kubota for her comments and suggestions that improved this work.
RM thank Fausto Vagnetti for useful discussions. This  work  is  based  on  observations obtained with: the NuSTAR mission,  a  project  led  by  the  California  Institute  of  Technology,  managed  by  the  Jet  Propulsion  Laboratory  and  funded  by  NASA; XMM-Newton,  an  ESA  science  mission  with  instruments  and  contributions  directly funded  by  ESA  Member  States  and  the  USA  (NASA).
This  research  has  made  use  of  data,  software  and/or  web tools obtained from NASA's High Energy Astrophysics Science  Archive  Research  Center  (HEASARC),  a  service  of Goddard  Space  Flight  Center  and  the  Smithsonian  Astrophysical  Observatory. Part of this work is based on archival data, software or online services provided by the Space Science Data Center - ASI. The research leading to these results has received funding from the European Union's Horizon 2020 Programme under the AHEAD project (grant agreement n. 654215). RM \& SB acknowledge financial support from ASI under grants ASI-INAF I/037/12/0 and n. 2017-14-H.O.
GM \& AM acknowledges financial support from the Italian Space Agency under grant ASI/INAF I/037/12/0-011/13.
POP thanks financial support from the CNES and the CNRS/PNHE.
FU acknowledges financial support from ASI under contract ASI/INAF 2013-025-R1. ADR acknowledges financial contribution from the agreement ASI-INAF n.2017-14-H.O.
BDM acknowledges support from the Polish National Science Center grant Polonez 2016/21/P/ST9/04025.




\bibliographystyle{mnras}
\bibliography{4593.bib} 








\bsp	
\label{lastpage}
\end{document}